\documentclass[11pt,aps,showpacs]{revtex4}
\usepackage{hyperref}
\usepackage{graphicx}
\usepackage{amssymb}
\usepackage{amsmath}
\usepackage{slashed}
\newcommand{\nn}{\nonumber \\}
\newcommand{\be}{\begin{equation}}
\newcommand{\ee}{\end{equation}}
\newcommand{\bea}{\begin{eqnarray}}
\newcommand{\ea}{\end{eqnarray}}

\newcommand{\R}{{\mathbb R}}
\newcommand{\tr}{{\rm tr}}

\begin{document}

%----------------------------------------------------------------------------
\title{Gross-Neveu Models, Nonlinear Dirac Equations, Surfaces and Strings}
\author{G\"ok\c ce Ba\c sar and Gerald~V.~Dunne}
\affiliation{Department  of Physics, University of Connecticut, Storrs, CT 06269}
%\maketitle
%--------------------------------------------------------------------------

\begin{abstract}
Recent studies of the thermodynamic phase diagrams of the Gross-Neveu model (${\rm GN}_2$), and its chiral cousin, the ${\rm NJL}_2$ model, have shown that there are phases with inhomogeneous crystalline condensates. These (static) condensates can be found analytically because the relevant Hartree-Fock and gap equations can be reduced to the nonlinear Schr\"odinger equation, whose deformations are  governed by the mKdV and AKNS integrable hierarchies, respectively. Recently, Thies  et al have shown that time-dependent Hartree-Fock solutions describing baryon scattering in the massless ${\rm GN}_2$ model satisfy the Sinh-Gordon equation, and can be mapped directly to classical string solutions in $AdS_3$.  Here we propose a geometric perspective for this result, based on the generalized Weierstrass spinor representation for the embedding of 2d surfaces into 3d spaces, which explains why these well-known integrable systems underlie these various Gross-Neveu gap equations, and why there should be a connection to classical string theory solutions. This geometric viewpoint may be useful for higher dimensional models, where the relevant integrable hierarchies include the Davey-Stewartson and Novikov-Veselov systems.

\end{abstract}


%\date{\today}

\maketitle
%\section{}
%\subsection{}

The Gross-Neveu model \cite{Gross:1974jv,Dashen:1975xh,Feinberg:2003qz} is a remarkable $1+1$ dimensional interacting fermionic model that shares some important features with quantum chromodynamics (QCD): it is asymptotically free, it has dynamical chiral symmetry breaking, and it has a limit of large number of flavors that behaves like the 't Hooft large $N_c$ limit of QCD. It has been well studied, but some surprising features have come to light only relatively recently. For example, at finite temperature and density, in the $N_f\to\infty$ limit, the phase diagram shows regions in which the system prefers to form a spatially inhomogeneous crystalline condensate. This is true both of the Gross-Neveu model, ${\rm GN}_2$, that has a discrete chiral symmetry \cite{Thies:2003kk,Thies:2006ti}, and of the ${\rm NJL}_2$ model that has a continuous chiral symmetry \cite{Schon:2000he,Basar:2009fg}, although the phase diagrams are very different. The inhomogeneous phase of the ${\rm GN}_2$ model has been verified on the lattice \cite{deForcrand:2006zz}. Similar 1d inhomogeneous condensates have been found in related models, also in higher-dimensions  \cite{Nickel:2008ng,glozman,Kojo:2010zz,Frolov:2010wn}. The relevant gap equations can be solved analytically at finite temperature and density because of a rich integrability structure underlying the Gross-Neveu models. Important aspects of this integrability were recognized already in the zero temperature and zero density analyses \cite{Dashen:1975xh,Pohlmeyer:1975nb,Lund:1976ze,Neveu:1977cr}, but the integrability properties have profound implications for the analysis of these models at nonzero temperature and nonzero density \cite{Basar:2008ki,Correa:2009xa}. For example, the thermodynamic Ginzburg-Landau expansions of the ${\rm GN}_2$ and ${\rm NJL}_2$ models are expansions in functionals of the associated condensate, and these functionals are precisely  the conserved quantities of the modified Korteweg-de Vries (mKdV) \cite{gesztesy} and Ablowitz-Kaup-Newell-Segur (AKNS) hierarchies \cite{Ablowitz:1974ry}, respectively; this permits the gap equation to be solved and re-summed to all orders. The (static) condensate satisfies the nonlinear Schr\"odinger equation, and the deformations of this equation encode the spectral information needed to study the thermodynamics; these deformations
are governed by the mKdV and AKNS hierarchies, respectively, for real or complex condensates.

More recently  \cite{Klotzek:2010gp}, it has been shown that there are real time-dependent condensates that satisfy the Hartree-Fock equations of the ${\rm GN}_2$ system, and that these condensates are associated with the Sinh-Gordon equation, in light-cone coordinates. These solutions describe various baryon scattering processes  \cite{Klotzek:2010gp,Fitzner:2010nv}; furthermore, they can be mapped directly  \cite{Klotzek:2010gp} to various classical string solutions in ${\rm AdS}_3$  \cite{Jevicki:2007aa}. This is a much deeper result than that concerning static inhomogeneous condensate solutions, and so it raises the question of how this works in the gap equation language, and how this relates to the static condensates. Furthermore, at first sight it is quite surprising to see such an explicit connection between the ${\rm GN}_2$ model and string theory. While relations between string theory and Gross-Neveu models have been proposed previously \cite{Antonyan:2006qy,Basu:2006eb,Davis:2007ka}, these have not been based on the inhomogeneous phases at finite temperature and density.

In this paper we suggest a natural geometric explanation for the origin of the integrable Sinh-Gordon system in the Hartree-Fock analysis of the ${\rm GN}_2$ model, and our explanation makes it immediately clear why there is such a direct connection with classical ${\rm AdS}_3$ string solutions. Our proposal is strongly motivated by the seminal work of Regge and Lund \cite{Lund:1976ze}, who first explicitly recognized the geometric connection between string solutions, the sine-Gordon equation and the problem of embedding surfaces, and the work of Pohlmeyer \cite{Pohlmeyer:1975nb}, and Neveu and Papanicolaou \cite{Neveu:1977cr}, connecting sigma models, integrable systems and fermionic models.
Our geometric explanation is based on the fact that the inhomogeneous condensates (static and time-dependent) of the ${\rm GN}_2$ model all come from spinors that satisfy a nonlinear Dirac equation (in the terminology of  \cite{Klotzek:2010gp,Fitzner:2010nv} they are "Type I": they solve the Hartree-Fock equations mode-by-mode, see (\ref{nld}) below). This nonlinear Dirac equation has a simple geometrical interpretation in terms of the embedding of surfaces of constant mean curvature into 3 dimensional spaces. The key mathematical result is simple to state: it is a classical result in differential geometry \cite{eisenhart-book,hopf-book,kenmotsu,konopelchenko1,bobenko-integrable,fokas,taimanov-dirac,Novikov:2006zz} that the problem of immersing a 2d surface into a higher dimensional space may be expressed naturally in spinor language, with the geometric Gauss-Codazzi equations appearing as the compatibility conditions for a Dirac equation. Initially suggested by Weierstrass and Enneper for minimal surfaces in $\mathbb R^3$, this construction is now recognized as being far more general, even describing embeddings of surfaces into  Lie groups and symmetric spaces.
For surfaces of {\it constant mean curvature}, this Dirac equation  becomes a {\it nonlinear} Dirac equation, and we show here that with appropriate choices of signature and manifolds, it is precisely the  nonlinear Dirac equation of the ${\rm GN}_2$ model. Furthermore, the  Sinh-Gordon equation appears because for constant mean curvature surfaces the square root of the induced metric factor of the emdedded surface is proportional to the spinor condensate, and the Gauss equation for this metric factor is precisely of Sinh-Gordon form.
The relation to classical string solutions follows because constant mean curvature surfaces can be viewed as minimal surfaces [i.e. zero mean curvature], which are just string solutions, immersed in a space of constant curvature, such as ${\rm AdS}_3$. The relation to the static solutions found in \cite{Thies:2006ti,Basar:2009fg,Basar:2008ki,Correa:2009xa} is that the dimensionally reduced version is the problem of immersing a curve [rather than a surface] into the 3d space, and this is also a classical problem in differential geometry, for which the familiar Frenet-Serret equations can be written in Dirac form, a fact that has been rediscovered many times in various physical and mathematical contexts, ranging from the motion of vortex filaments to optical solitons \cite{darios,lamb,hasimoto,Sym:1979er,dodd,calini,schmidt}.

In Section I we review the Hartree-Fock and gap equation approaches to the Gross-Neveu model ${\rm GN}_2$. Section II describes the surface embedding problem for three different cases: (i) embedding a Euclidean signature 2d surface into $\mathbb R^3$; (ii) embedding a Minkowski signature 2d surface into $\mathbb R^{1,2}$; (iii) embedding a Minkowski signature 2d surface into ${\rm AdS}_3$. The first is the simplest case, which we review as an introduction,  while we show that the Minkowski cases are the ones that are relevant for the Gross-Neveu model ${\rm GN}_2$. In Section III we describe the reduction to the previously known static inhomogeneous condensates. In Section IV we show how these ideas can be used to search for inhomogeneous condensates in the $2+1$ dimensional Gross-Neveu model, a model for which much less is known.

\section{Gross-Neveu model, Hartree-Fock and Gap Equations}

The Gross-Neveu model (${\rm GN}_2$) is a massless self-interacting fermionic theory in two dimensional spacetime with Lagrangian \cite{Gross:1974jv}:
\begin{eqnarray}
{\mathcal L}_{\rm GN}&=&i\,\bar{\psi}\,\partial\hskip -6pt /\,\psi +\frac{g^2}{2}\, (\bar \psi\,\psi)^2 
%{\mathcal L}_{\rm NJL}&=&i\,\bar{\psi}\,\partial\hskip -6pt /\,\psi +g\left[(\bar \psi\,\psi)^2+(\bar \psi\,i\, \gamma^5\psi)^2\right]
\label{lag}
\end{eqnarray}
where $g^2$ is a coupling constant. This model is interesting because it is asymptotically free, it exhibits dynamical mass generation, and it has a (discrete) chiral symmetry that is spontaneously broken. We have suppressed the flavor indices as we work exclusively in the ``'t Hooft limit'' of $N_f\to\infty$, with $g^2N_f$ fixed.
The Dirac operator is $\partial\hskip -6pt /=\gamma^\mu\partial_\mu$, with Dirac matrices $\gamma^0$ and $\gamma^1$,
to be specified later (for the Euclidean and Minkowski cases).
%, and $\gamma^5=\gamma^0\gamma^1$.
Note that we use the relativistic physics notation, $\bar\psi\equiv \psi^\dagger\, \gamma^0$. [A word of caution: the mathematical literature concerning the spinor representation of surface embeddings often uses $\bar\psi$ to mean $\psi^\dagger$. The distinction is very important.]

There are two equivalent methods to seek approximate solutions for the Gross-Neveu model. One method, the Hartree-Fock approach, is to seek self-consistent solutions to the Dirac equation
\begin{eqnarray}
\left(i\,\partial\hskip -6pt /\, -S\right)\psi&=&0
%\qquad , \qquad \langle \bar\psi\, \psi\rangle=S 
%{\rm NJL}: \left(i\,\partial\hskip -6pt /\,\psi -S-iP\gamma^5\right)\psi&=&0\qquad , \qquad \langle \bar\psi\, \psi\rangle-i \langle \bar\psi\,i\, \gamma^5 \psi\rangle=S-iP
\label{hf}
\end{eqnarray}
subject to the constraint
\begin{eqnarray}
\langle \bar\psi\, \psi\rangle\equiv \sum_{{\rm states}\, p} \, \bar\psi_p\, \psi_p = -\frac{1}{g^2}\, S
%\qquad , \qquad \langle \bar\psi\,i\, \gamma^5\, \psi\rangle=\sum_{{\rm states}\, k} \, \bar\psi_k\,i\, \gamma^5\, \psi_k
\label{exp}
\end{eqnarray}
The simplest way [but not necessarily the only way, as we discuss later]  to solve these equations is to find  solutions $\psi_p$ to the nonlinear Dirac equation {\it for each mode $p$ separately}:
\begin{eqnarray}
\left(i\,\partial\hskip -6pt /\, -l\, (\bar\psi_p\,\psi_p)\right)\psi_p&=&0 
%\cr{\rm NJL}: \left(i\,\partial\hskip -6pt /\,\psi -g(\bar\psi_k\,\psi_k)-i\,g(\bar\psi_k\,i\,\gamma^5\,\psi_k)\gamma^5\right)\psi_k&=&0
\label{nld}
\end{eqnarray}
for some constant $l$, whose physical significance will become clear below. Thies et al refer to these mode-by-mode solutions to the nonlinear Dirac equation (\ref{nld}) as ``Type I'' solutions  \cite{Klotzek:2010gp,Fitzner:2010nv}. All known static and time-dependent solutions of the Hartree-Fock equations for ${\rm GN}_2$ are of this ``Type I'' form.
The temperature ($T$) chemical potential ($\mu$) phase diagram of the massless ${\rm GN}_2$ model is characterized by a {\it static} crystalline condensate which has the form of a kink crystal:
\begin{eqnarray}
S(x)=m\, k\,{\rm sn}(m\, x; k^2)
\label{sn}
\end{eqnarray}
where sn is the usual Jacobi elliptic function, with elliptic parameter $k^2$. With this $S(x)$, the corresponding solutions $\psi_p$ to the nonlinear Dirac equation can be constructed analytically, and the Hartree-Fock equations (\ref{hf}) and (\ref{exp}) are satisfied \cite{Thies:2003kk,Thies:2006ti}.
This kink crystal condensate (\ref{sn}) is characterized by two parameters, $m$ and $k^2$, which together determine the amplitude, period and shape of the crystalline condensate. Thermodynamic minimization of the associated grand potential determines $m$ and $k^2$ as functions of $T$ and $\mu$, so that the form of the crystal changes depending on temperature and density \cite{Thies:2003kk,Thies:2006ti,Basar:2009fg}. Indeed in one region of the phase diagram a vanishing condensate is thermodynamically preferred ($m=0$), while in another region a constant homogeneous condensate is preferred ($m\neq 0$, and $k^2=1$), while in the crystalline phase the preferred condensate has $m\neq 0$, and $k^2<1$.

There are also exact {\bf time-dependent} ``Type I'' solutions $S(x, t)$ of the time-dependent Hartree-Fock equations, (\ref{hf}) and (\ref{exp}), that also satisfy the nonlinear Dirac equation (\ref{nld}) mode-by-mode \cite{Klotzek:2010gp}. The simplest nontrivial example is the boosted kink 
\bea
S(x, t)=\tanh(2(x-v t)/\sqrt{1-v^2})
\label{bk}
\ea
from which one can construct a real kink-antikink scattering solution:
\begin{eqnarray}
S(x, t)=\frac{v\, \cosh(2x/\sqrt{1-v^2})-\cosh(2v t/\sqrt{1-v^2})}{v\, \cosh(2x/\sqrt{1-v^2})+\cosh(2v t/\sqrt{1-v^2})}
\label{kk}
\end{eqnarray}
The type I breather solution is not real \cite{Klotzek:2010gp}.
Using the relation to the Sinh-Gordon equation, one can use B\"acklund transformations to construct more complicated solutions describing multi-baryon scattering processes \cite{Klotzek:2010gp,Fitzner:2010nv}. Again, these time-dependent $S(x, t)$ satisfy the time-dependent Hartree-Fock problem, and the corresponding solutions $\psi_p$ to (\ref{hf}) can be constructed analytically, and satisfy the nonlinear Dirac equation (\ref{nld}) for each mode $p$.
 
 A second, equivalent, method to approach the ${\rm GN}_2$ model is via a  functional gap equation, where one replaces the fermion bilinear $\bar\psi\psi$  by a bosonic condensate, $S$, leading to a gap equation (we suppress the coupling $g$):
 \begin{eqnarray}
% &&{\rm GN}: 
S(x,t)=\frac{\delta}{\delta S(x,t)}\,\ln\det \left(i\,\partial\hskip -6pt /- S(x,t) \right) 
%&&{\rm NJL}: \Delta=\frac{\delta}{\delta \Delta^*}\,\ln \det \left(i\,\partial\hskip -6pt /\, -g(S-iP\gamma^5)\right)
\label{gap}
\end{eqnarray}
When $S$ is constant, it is straightforward to solve this gap equation \cite{wolff,treml,barducci}. However, such a solution does not give the correct phase diagram  \cite{Thies:2003kk,Thies:2006ti,Basar:2009fg}.
When the condensate $S$ is inhomogeneous, $S=S(x,t)$, it is a highly nontrivial matter to solve the functional differential equation (\ref{gap}). For static, but spatially inhomogeneous, condensates $S(x)$ it has been shown \cite{Basar:2008ki} that the gap equation can be mapped to the solution of the nonlinear Sch\"odinger equation (NLSE), 
\bea
S^{\prime\prime}-2 S^3+c\, S =0
\label{nlse}
\ea
where $c$ is some constant. The most general bounded solution of (\ref{nlse}) is  given by the kink crystal  (\ref{sn}), with the identitifcation: $c=m^2(1+k^2)$. Thermodynamic minimization of the corresponding grand potential requires studying this solution as the two parameters $m$ and $k^2$ are varied, which means we study deformations of the NLSE. These deformations are integrable and are governed by the mKdV hierarchy \cite{Basar:2008ki,Correa:2009xa}, which means that the gap equation (\ref{gap}) is precisely the Novikov equation for mKdV. More explicitly, in thermodynamic language, when $S=S(x)$, we can expand the log det in a Ginzburg-Landau expansion:
\begin{eqnarray}
\ln\det \left(i\,\partial\hskip -6pt /-S(x) \right) &=&\sum_{n} \alpha_{n}(T, \mu)\, a_n[S]
%{\rm NJL}:\qquad \ln \det \left(i\,\partial\hskip -6pt /\, -g(S-iP\gamma^5)\right) &=& \sum_n \alpha_n(T, \mu)\, b_n[S, P]
\label{gl}
\end{eqnarray}
where $\alpha_n(T, \mu)$ are simple known functions of the temperature $T$ and chemical potential $\mu$, and $a_n[S]$ are known functionals of the condensate $S(x)$ and its spatial derivatives \cite{Basar:2008ki,Correa:2009xa}. Indeed, the $a_n[S]$ are precisely the conserved quantities of the mKdV hierarchy. Using the integrability properties of the mKdV hierarchy, one finds that the solutions $S(x)$ of the NLSE satisfy the gap equation (\ref{gap}) {\it term-by-term}, in the sense that
\begin{eqnarray}
\frac{\delta a_n[S]}{\delta S(x)}\propto  S(x)\qquad, \qquad {\rm for\,\, all\,\,} n
\label{gap-mode}
\end{eqnarray}
This is the gap equation analogue  \cite{Basar:2008ki}  of the mode-by-mode solution of the Hartree-Fock problem in (\ref{nld}).

In this gap equation language, the observation of Thies et al \cite{Klotzek:2010gp,Fitzner:2010nv}, concerning the time-dependent solutions of the Hartree-Fock problem, is truly remarkable: it means that one has an analytic {\it space- and time-dependent} solution $S(x,t)$ to the functional gap equation (\ref{gap}). At first sight, it is also surprising that these solutions should be directly related to classical string solutions. The connection to string solutions pointed out in \cite{Klotzek:2010gp,Fitzner:2010nv} is via the Sinh-Gordon equation, which permits use of the explicit solutions constructed by Jevicki et al \cite{Jevicki:2007aa}. For these time-dependent condensates, $S(x,t)$, the role of the NLSE for static solutions is played by the Sinh-Gordon equation. This has motivated us to look for the most natural language in which to understand  these time-dependent solutions. We suggest here that the essence of the relation is due to an elementary property of differential geometry:  the immersion of two dimensional surfaces is naturally expressed in terms of spinors and a Dirac-like equation, and for surfaces of constant mean curvature, this Dirac equation becomes a nonlinear Dirac equation. To match precisely to the ${\rm GN}_2$ model we have to choose carefully the surface and the space into which it is immersed, but the basic idea is due to this generalized Weierstrass representation of surface immersions, a classical result of differential geometry.

\section{Nonlinear Dirac Equation and Surface Embeddings}

In this section we relate the nonlinear Dirac equation (\ref{nld}) to the problem of immersion of a 2d surface into a 3d space \cite{eisenhart-book,hopf-book}. We begin with a review of the Euclidean signature case, where the most important result for our purposes is the generalized Weierstrass representation of a surface \cite{kenmotsu,bobenko-integrable,konopelchenko1,fokas,taimanov-dirac,Novikov:2006zz}, which states that the solution to the 2d (Euclidean) Dirac equation
\begin{eqnarray}
i\left(\partial \hskip -6pt /\hskip 3pt -S\right)\psi=0
\label{dirac}
\end{eqnarray}
defines a 2d surface in ${\mathbb R}^3$, such that the mean curvature $H$ [defined below]  of the surface is related to the potential $S$ appearing in the Dirac equation (\ref{dirac}) as:
\begin{eqnarray}
H=\frac{S}{\psi^\dagger\psi}
\label{hs}
\end{eqnarray}
Thus, for a surface of constant mean curvature, $H=l$, the Dirac equation (\ref{dirac}) becomes a nonlinear Dirac equation:
\begin{eqnarray}
i\left(\partial \hskip -6pt /\hskip 3pt -l\, \psi^\dagger\psi\right)\psi=0
\label{nldirac}
\end{eqnarray}
Notice that while this is a nonlinear Dirac equation, it is not  of the Gross-Neveu form (\ref{nld}), where the nonlinear term is $\bar\psi\psi$, instead of $\psi^\dagger \psi$. We then show that the nonlinear term $\bar\psi\psi$ arises from embedding a Minkowski signature surface into a 3d space such as ${\mathbb R}^{1,2}$ or $AdS_3$. To establish notation and to compare with classical differential geometry, we first quickly review the standard Euclidean embedding case, the generalized Weierstrass representation, before turning to the Minkowski cases.

\subsection{Euclidean Signature: Conformal Immersion of 2d Surfaces in ${\mathbb R}^3$}

\subsubsection{Basic Differential Geometry of Euclidean Surface Embedding}

\begin{figure}[htb]
\includegraphics[scale=0.25]{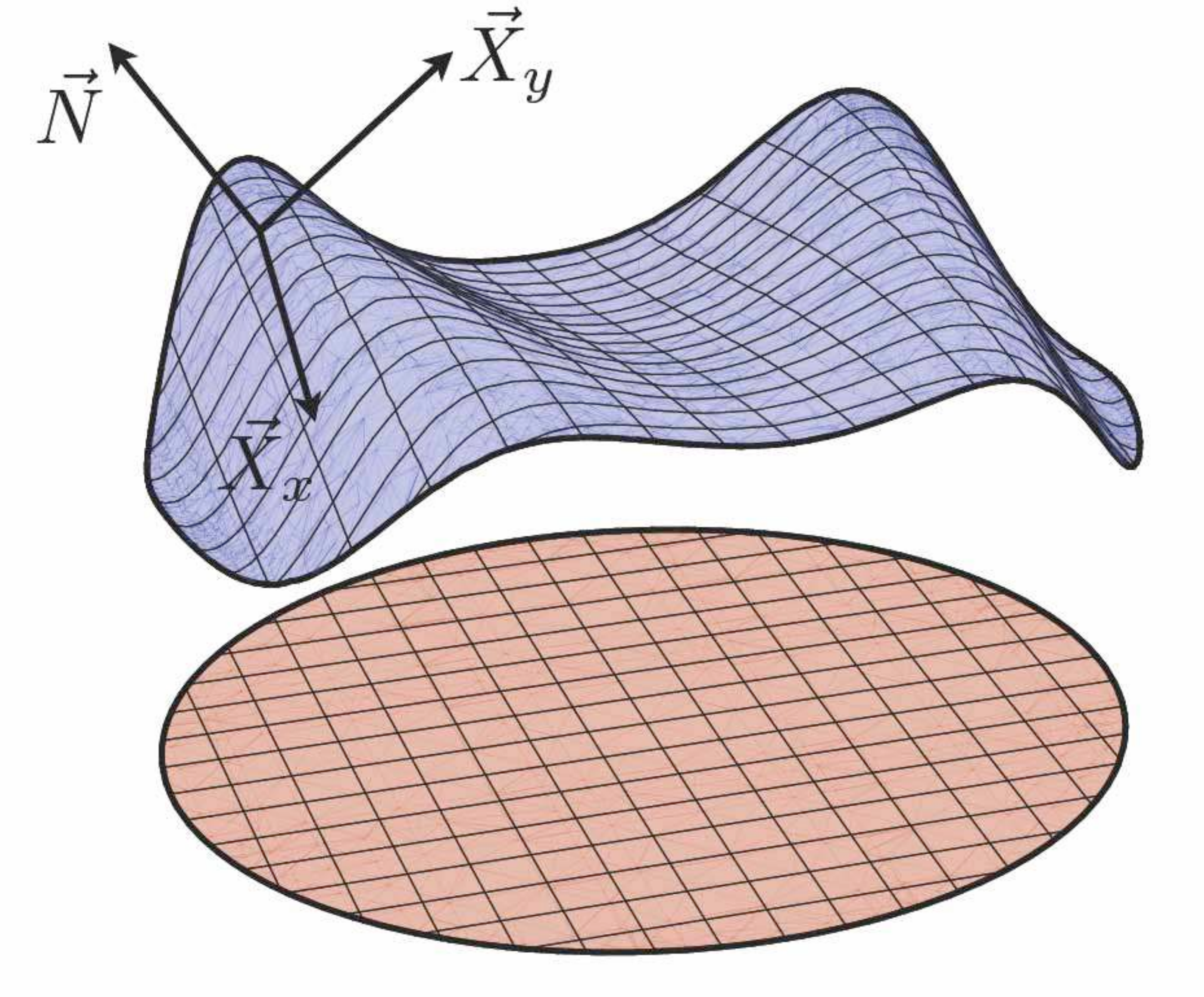}
\caption{Embedding of a 2d surface into $\mathbb R^3$. The surface can be characterized by the motion of the frame $\{\vec N, \vec X_x, \vec X_y\}$, consisting of the normal and tangent vectors, as it moves across the surface. This is described by the Gauss-Codazzi equations.}
\end{figure}
Consider a smooth, orientable surface $\vec{X}$, parametrized by $(x,y)\in \Sigma\subset \R^2$, embedded conformally into $\R^3$ \cite{eisenhart-book,hopf-book}:
\bea
\vec X(x,y): \Sigma \mapsto \R^3
\ea
The conformal parametrization fixes the induced metric of the surface to be of the form $ds^2=f^2(x,y)(dx^2+dy^2)$, and in complex coordinates, $z=x+iy$, $\bar z=x-iy$, we have
\bea
ds^2=f^2(z,\bar z)\,dz\,d\bar z
\label{complex_metric}
\ea 
The isometries of the metric (\ref{complex_metric}) are holomorphic mappings $z\to\omega(z)$.
The surface can be described by the frame of three orthogonal vectors: the unit normal $\vec N$ to the surface, and the two derivatives, $\vec X_x\equiv \partial_x \vec X$, and $\vec X_y\equiv \partial_y \vec X$. It is more convenient to consider the complex tangent vectors, $\vec X_z\equiv \partial_z \vec X$, and $\vec X_{\bar z}\equiv \partial_{\bar z} \vec X$, together with the the unit normal vector $\vec N$. These three vectors form a basis $\{\vec e_{(i)}\}=\{ \vec X_z ,\vec X_{\bar z}, \vec N \}$ in $\mathbb C^3$, with inner products:
\bea
\vec X_z.\vec X_{\bar z}=\frac{f^2}{2}\quad,\quad\vec X_z.\vec X_z=0\quad,\quad\vec X_{\bar z}.\vec X_{\bar z}=0
\label{complex_cond}
\ea 
These inner products define the elements of the "first fundamental form".
%\bea
%\vec X_z.\vec X_{\bar z}=\frac{f^2}{2}\quad,\quad\vec X_z.\vec X_z=0\quad,\quad\vec X_{\bar z}.\vec X_{\bar z}=0
%\label{complex_cond}
%\ea 
%Also 
To characterize the surface, we consider how the frame changes as we move across the surface, which requires the second derivatives.
In order to analyze the evolution of this moving frame, we expand $\vec X_{zz}$, $\vec X_{\bar z\bar z}$ and $\vec X_{z\bar z}$, in the basis $\{ \vec X_z ,\vec X_{\bar z}, \vec N \}$. The relevant basis projections are contained in the inner products
\bea
\vec X_{zz}.\vec N=Q\quad,\quad
\vec X_{\bar z \bar z}.\vec N=Q^*\quad,\quad 
\vec X_{z\bar z}.\vec N=\frac{1}{2}f^2 H 
\ea
which define the {\it Hopf differential}, $Qdz^2$, and the {\it mean curvature}, $H$. The condition of commutativity of the derivatives, $\vec e_{(i),\,z\bar z}=\vec e_{(i),\,\bar z z}$, generates the Gauss-Codazzi equations:
\bea
f f_{z \bar z}-f_z f_{\bar z} +\frac{1}{4} H^2 f^4&=& |Q|^2 
\label{gauss-e}\\
Q_{\bar z}&=&\frac{1}{2} f^2 H_z
\label{codazzi-e1}\\
Q_{z}^*&=&\frac{1}{2} f^2 H_{\bar z}
\label{codazzi-e2}
\ea
It is a simple exercise to show that these equations are invariant under isometries.

Functions $f,Q$ and $H$ satisfying the Gauss-Codazzi equations (\ref{gauss-e}, \ref{codazzi-e1}, \ref{codazzi-e2}) are necessary and sufficient to characterize a surface $\vec X$ in $\R^3$ \cite{eisenhart-book,hopf-book}. The associated Gaussian curvature, $K$, which is an intrinsic quantity, is given by 
%We can easily be calculate the gaussian curvature through the first and second fundamental forms: 
%\bea
%I=f^2 \begin{pmatrix} 1\, &  0 \cr  0\, & 1 \end{pmatrix} \quad II= \begin{pmatrix} 2\,\mathcal ReQ + H f^2  &  -2 \,\mathcal ImQ  \cr -2\,\mathcal Im Q  &  -2\,\mathcal Re Q + H f^2 \end{pmatrix} 
%\nn \nn
%K=det(I^{-1}II)=H^2-\frac{4|Q|^2}{f^4}=-2f^{-2}\ln(f^2)_{z\bar z}
%\ea
\bea
K=H^2-\frac{4|Q|^2}{f^4}=-2f^{-2}\ln(f^2)_{z\bar z}
\label{k}
\ea
where in the last step we used the Gauss-Codazzi equation (\ref{gauss-e}). 
Two important well-studied cases are surfaces of constant Gaussian curvature,  and surfaces of constant mean curvature. In the first case, with $K$=constant, we see from ({\ref{k}) that the metric must satisfy the Liouville equation
\bea
\ln(f^2)_{z\bar z}=-2Kf^2
\label{liouville}
\ea
In the second case, with  $H$  constant, $H=l$,  we note that 
%Two important special surfaces are the ones with constant gaussian curvature and constant mean curvature. The former is described by to the Liouville's equation. Depending on the sign of $K$ these are isomorphic to sphere ($K>0$), plane ($K=0$) or a pseudosphere ($K<0$). 
the Gauss-Codazzi equations (\ref{codazzi-e1}, \ref{codazzi-e2}) imply that the Hopf differential is holomorphic, $Q=Q(z)$, and therefore  there exists an isometry where it is constant. Then, writing the metric factor as $f^2=\frac{2|Q|}{H}e^\theta$, the first of the Gauss-Codazzi equations (\ref{gauss-e}) becomes the Sinh-Gordon (ShG) equation:
\bea
\theta_{z\bar z}+2\,l\, |Q|\, \sinh(\theta)=0
\label{shg}
\ea

\subsubsection{Spinor Representation of Euclidean Surfaces}

The generalized Weierstrass representation  expresses the Gauss-Codazzi equations in terms of a Dirac equation, and an associated zero-curvature condition.
We choose the Euclidean gamma matrices, $\gamma_0=-i\sigma_1$, and $\gamma_1=-i\sigma_2$. Writing $\psi=(\psi_1, \psi_2)^T$, the Euclidean Dirac equation (\ref{dirac}) becomes
\bea
\begin{pmatrix}
S& -2 \partial_z\cr
2\partial_{\bar z} &S
\end{pmatrix}
 \psi=0\qquad \Leftrightarrow \qquad
\begin{cases}
\psi_{1, \bar z}= -\frac{1}{2}S\,\psi_2\cr
 \psi_{2, z} = \frac{1}{2} S\,\psi_1
 \end{cases}
\label{dirac-e2}
\ea
Note that if $\psi=(\psi_1,\psi_2)^T$ is a solution, and $S$ is real, then $(-\psi^*_2,\psi^*_1)^T$ is also a solution. From these two solutions, form the $2\times 2$ matrix $\Phi$:
\bea
\Phi=\left(\begin{matrix} \psi_1 & -\psi^*_2 \\  \psi_2\,\, & \psi^*_1 \end{matrix} \right)
\label{phi-e}
\ea      
for which
\bea
\det \Phi=\psi^\dagger \psi
\label{det-e}
\ea
The connection to surface embeddings is expressed most compactly using the identification between vectors $\vec X \in \R^3$ and quaternions $X\in {\mathbb H}$:
\bea
%&{\bf\hat i}\equiv-i\sigma_1=-i\left(\begin{matrix} 0\, & 1\\ 1\, & 0 \end{matrix} \right) \quad {\bf\hat j}\equiv-i\sigma_2=-i\left(\begin{matrix} 0\, & -i \\  i\, & 0 \end{matrix} \right) \quad {\bf\hat k}\equiv-i\sigma_3=-i \left(\begin{matrix} 1\, & 0 \\  0\, & -1 \end{matrix} \right)& \\
\vec X=(X_1,X_2,X_3)\quad &\leftrightarrow& \quad X=-i
\begin{pmatrix}
X_3 & X_1-i X_2\cr
X_1+i X_2 &-X_3
\end{pmatrix} 
\\
X_a &=&\frac{i}{2}\tr\left(X\, \sigma_a\right)
\ea 
where $\sigma_a$ are the Pauli matrices. Now, given the solution $\Phi$ in (\ref{phi-e}), we construct the quaternionic representation of three vectors as
\bea
X_z=-if\,\Phi^{-1}\left(\begin{matrix} 0\, & 0 \\  1\, & 0 \end{matrix} \right)\Phi 
\quad , \quad 
X_{\bar z}=-if\,\Phi^{-1}\left(\begin{matrix} 0\, & 1 \\  0\, & 0 \end{matrix} \right)\Phi 
\quad,\quad 
N=-i\,\Phi^{-1}\sigma_3\Phi
\ea
where $f=\det \Phi$. 
It is straightforward to check that these define the moving frame of a surface, $\vec X=\frac{i}{2}\tr\left(X\, \vec\sigma\right)$, and the geometric
%Gauss-Weingarten 
equations for the moving frame appear as the linear matrix equations
\bea
\Phi_z=U\,\Phi \qquad,\qquad \Phi_{\bar z}=V\,\Phi
\label{UV}
\ea
 where $U$ and $V$ are $2\times2$ matrices. Using $X_{z\bar z}=X_{\bar z z}$ we can express them in terms of the geometric quantities:
 \bea
U=\begin{pmatrix} 
\frac{f_z}{f}\, & -\frac{Q}{f}  \cr  
\frac{H}{2}f\,\, & 0 
\end{pmatrix}  
\qquad,\qquad 
V=\begin{pmatrix} 
0\, & -\frac{H}{2}f \cr
 \frac{\bar Q}{f} \, & \frac{f_{\bar z}}{f} 
 \end{pmatrix} 
\label{UV_geom_eucl}
\ea
%[Note that the normalization of $\Phi$ fixes $\tr\,U=\frac{f_z}{f}$ and $\tr\,V=\frac{f_{\bar z}}{f}$.]
By construction, the compatibility of the two linear equations in (\ref{UV}) implies that $U$ and $V$ satisfy the zero curvature  condition, 
\bea
U_{\bar z}-V_z+[U,V]=0
\label{eucl_lax}
\ea
and this zero curvature condition encodes the Gauss-Codazzi equations (\ref{gauss-e}, \ref{codazzi-e1}, \ref{codazzi-e2}). 

This construction is known as the generalized  Weierstrass representation of the embedded surface, and all the geometrical information can be encoded in the Dirac equation (\ref{dirac-e2}). 
To summarize, we establish a dictionary between the Euclidean surfaces in $\R^3$ and spinors:
\bea
\text{Dirac equation}&:&\, i\left(\partial \hskip -6pt /\hskip 3pt -S\right)\psi=0\nn
\text{Mean curvature}&:&\,H=S\,(\psi^\dagger\psi)^{-1}\nn
\text{induced metric factor}&:&\, f=\det\Phi=\psi^\dagger\psi \nn
\text{Hopf differential}&:&\,\begin{cases}
Q=(\psi^*_{2,z}\,\psi_1-\psi^*_2\,\psi_{1,z})\cr
Q^*=(\psi_{2,\bar z}\,\psi_1^*-\psi_2\,\psi_{1,\bar z}^*)
\end{cases}
\label{dic_eucl}
\ea
Minimal surfaces (with $H=0$) are described by a \textit{free massless} Dirac equation, with $S=0$, and the corresponding spinors are (anti)holomorphic. This is the original Weierstrass-Enneper representation. The more general constant mean curvature surfaces (with $H=l$) are described  by a \textit{nonlinear} Dirac equation, since $H=l$ implies $S= l\, \psi^\dagger\psi$:
\bea
\text{Euclidean constant mean curvature surfaces}\quad\leftrightarrow\quad i(\slashed\partial-l\, \psi^\dagger\psi)\psi=0
\ea
Furthermore, the Gauss equation (\ref{gauss-e}) becomes the ShG equation
\bea
\text{Gauss equation}\quad \leftrightarrow\quad \text{Sinh-Gordon equation}:\quad
\theta_{z\bar z}+2 |Q| \, l\,\sinh(\theta)=0
\ea
where the metric factor $f^2$ is written as $f^2=\frac{2|Q|}{l}\, e^\theta$. In terms of the function $S(z, \bar z)$ appearing in the Dirac equation, this ShG equation reads:
\bea
S\, S_{z \bar z}-S_z\, S_{\bar z}+\frac{S^4}{4}=l^2 |Q|^2
\label{seq-e}
\ea
Note that the general solutions of the ShG equation, and also the corresponding solutions of the Dirac equation, can be expressed in terms of theta functions \cite{bobenko-integrable}.

\subsection{Minkowski Signature:  Immersion of 2d Surfaces in ${\mathbb R}^{1,2}$}

\subsubsection{Basic Differential Geometry of Minkowski Surface Embedding}

To study the Gross-Neveu model in two-dimensional Minkowski spacetime it is more appropriate to use the light-cone coordinates, $x_{\pm}=x \pm t$, instead of the complex coordinates, $z=x+i y$, $\bar z=x-i y$. 
So we now consider  an embedding of the form $\vec X(x,t): \Sigma \subset \R^{1,1} \mapsto \R^{1,2}$, with the conformal parameterization $ds^2=f^2(-dt^2+dx^2)$. The first fundamental form in light-cone coordinates is given by the inner products:
\bea
\vec X_+.\vec X_-=\frac{f^2}{2}\quad,\quad\vec X_+.\vec X_+=0\quad,\quad\vec X_-.\vec X_-=0
\label{first_lc}
\ea 
To construct the Gauss-Codazzi equations, we define the Minkowski analogues of the Hopf differential and the mean curvature, from the projections of the second derivatives:
\bea
\vec X_{++}.\vec N=-i\,Q^{(+)}\quad,\quad\vec X_{--}.\vec N=-i\,Q^{(-)}\quad,\quad\vec X_{+-}.\vec N=-\frac{i}{2}H\,f^2
%\vec X_{++}.\vec N=-i\,h_1\quad,\quad\vec X_{--}.N=-i\;h_2\quad,\quad\vec X_{+-}.N=-\frac{i}{2}H\,f^2
\ea
%The corresponding second fundamental form, $\Omega_{II}$, and 
The Gauss-Codazzi equations can easily be written by repeating the steps of the Euclidean construction in Minkowsi space: 
%\footnote{At the first sight the overall factor of $i$  seems a little unnatural;  however, it appears for the following matters of convention: in the Gross-Neveu language it is more natural to have a local Minkowski metric $(dt^2-dx^2)$, in order for the Dirac equation to have a real, rather than imaginary,  potential $S$. On the other hand, we wish eventually to match our construction to classical string solutions on $AdS_3$, for which the local surface metric of signature $(-dt^2+dx^2)$ is more natural.}
%\bea
%\Omega_{II}&=& i\left( \begin{matrix}   H f^2+h_1-h_2  &  h_1+h_2  \\  h_1+h_2  &   -H f^2+h_1-h_2 \end{matrix} \right) \nn\nn
%K&=&-H^2-\frac{4h_1\,h_2}{f^4}
%\ea
%The Gauss-Codazzi equations are
\bea
f f_{+-}-f_+ f_{-} -\frac{1}{4} H^2 f^4&=&-Q^{(+)}Q^{(-)}
%f f_{+-}-f_+ f_{-} -\frac{1}{4} H^2 f^4&=&h_1\,h_2 
\label{gauss-m}\\
Q^{(+)}_{-}&=& \frac{1}{2} f^2 H_+ 
%h_{1,-}&=& -\frac{1}{2} f^2 H_+ 
\label{codazzi1-m}\\
Q^{(-)}_{+}&=&\frac{1}{2} f^2 H_-
%h_{2,+}&=&\frac{1}{2} f^2 H_-
\label{codazzi2-m}
\ea
%(Note that in the Euclidean case the Codazzi equation (\ref{codazzi-e}) is really two equations, one for the real and imaginary parts of the Hopf differential $Q$).
As in the Euclidean case, surfaces of constant mean curvature ($H=l$) are characterized by a Sinh-Gordon equation.  If $H$ is constant, $H=l$, then we see that the  Gauss-Codazzi equations (\ref{codazzi1-m}, \ref{codazzi2-m}) imply that $Q^{(\pm)}$ 
%$h_1$ and $h_2$  
are left/right moving. Furthermore, they can be set to be constants, by appropriate isometries.
% that have the form $x_{\pm}=x_{\pm}(\tilde x_{\pm})$. 
%\bea
%f f_{+-}-f_+ f_--\frac{1}{4} l^2 f^4=h_1\,h_2\equiv \text{constant}
%\label{sgf-lc}
%\ea
Defining $f^2=\frac{2\sqrt{|Q^{(+)}Q^{(-)} |}}{l}\, e^\theta$,
%$f^2=\frac{2\sqrt{|h_1h_2 |}}{l}\, e^\theta$, 
the other Gauss-Codazzi equation (\ref{gauss-m}) becomes the 
%$f=\frac{(-4h_1 h_2 l^2)^{\frac{1}{4}}}{l}e^{\frac{\theta}{2}}$ 
 Sinh-Gordon (ShG) equation [note the opposite sign compared to (\ref{shg})]:
 \bea
\theta_{+-}-2l\sqrt{|Q^{(+)}Q^{(-)}|}\,\sinh(\theta)=0
%\theta_{+-}-2 \sqrt{|h_1h_2 |}\, l\,\sinh(\theta)=0
\label{sg-lc}
\ea

\subsubsection{Spinor Representation of Minkowski Surfaces in ${\mathbb R}^{1,2}$}

The Minkowski version of the generalized Weierstrass representation follows analogously. But switching to Minkowski signature, the nonlinear Dirac equation of the Gross-Neveu model emerges in the constant mean curvature case. 

We define Minkowski Dirac matrices, $\gamma_0=\sigma_1$, and $\gamma_1=i\sigma_2$, and the Dirac equation is
\begin{eqnarray}
\left(i\partial \hskip -6pt /\hskip 3pt -S\right)\psi=0 \qquad \text{or} \qquad
\begin{pmatrix}
S & -2i\partial_+\cr
2i\partial_- & S
\end{pmatrix}
\psi=0
\label{dirac-m}
\end{eqnarray}
which we can write as two equations:
\bea
\psi_{1, -}=\frac{i}{2} S\,\psi_2\qquad , \qquad \psi_{2, +}=-\frac{i}{2}S\, \psi_1
\label{dirac-m2}
\ea
Now note that if $\psi=(\psi_1,\psi_2)^T$ is a solution of (\ref{dirac-m}), and $S$ is real, then  $\psi=(-\psi^*_1,\psi^*_2)^T$ is also a solution of (\ref{dirac-m}). Thus, we now form the $2\times 2$ matrix:
\bea
\Phi=\left(\begin{matrix} \psi_1 & -\psi^*_1 \\  \psi_2\,\, & \psi^*_2 \end{matrix} \right)
\label{phi-m}
\ea  
for which
\bea
\det\Phi=\bar \psi \psi
\label{det-m}
\ea  
These expressions should be contrasted with the Euclidean versions (\ref{phi-e}) and (\ref{det-e}).
The crucial difference from the Euclidean case is 
that the metric factor $f$ will be identified with the condensate $\bar\psi\psi$, instead of with the density $\psi^\dagger\psi$. Ultimately this is due to the change of symmetry from $SU(2)$ to $SU(1,1)$. The underlying symmetry is now $SU(1,1)$, and the basis vectors are parametrized by $2\times2$ matrices $\Phi$, together with the inner product $X\cdot Y=-\frac{1}{2}\tr(XY)$:
\bea
&X_+=if\,\Phi^{-1}\begin{pmatrix} 
0\, & 0 \cr  1\, & 0 
\end{pmatrix} \Phi 
\quad , \quad 
X_-=if\,\Phi^{-1}
\begin{pmatrix} 
0\, & 1 \cr  0\, & 0 
\end{pmatrix} \Phi 
\quad , \quad N=i\,\Phi^{-1}\sigma_3\Phi&\nn
&X_{++}\cdot N=-i Q^{(+)}
\quad,\quad  X_{--}\cdot N=-i Q^{(-)} 
\quad,\quad
 X_{+-}\cdot N=-\frac{i}{2}H\,f^2&
\ea
We  use the same normalization condition, $\det\Phi=f$. The motion of the basis frame is characterized by linear equations involving matrices $U$ and $V$:
\bea
\Phi_+=U\,\Phi\qquad,\qquad \Phi_-=V\,\Phi
\label{UV_lc}
\ea
In terms of  geometric quantities they have similar forms to the Euclidean case:
\bea
U=\begin{pmatrix} 
\frac{f_+}{f}\, & i\frac{Q^{(+)}}{f}  \cr  -i\frac{H}{2}f\,\, & 0 
\end{pmatrix}
 \qquad,\qquad 
 V=\begin{pmatrix} 
 0\, & i\frac{H}{2}f \cr  -i\frac{Q^{(-)}}{f} \, & \frac{f_-}{f} 
 \end{pmatrix} 
\label{UV_geom_lc}
\ea
The Gauss-Codazzi equations (\ref{gauss-m}, \ref{codazzi1-m}, \ref{codazzi2-m}) are encoded by the zero-curvature condition:
\bea
U_--V_++[U,V]=0
\label{zc-m}
\ea

%The Dirac equation follows from the zero curvature condition $U_-+V_++[U,V]=0$:
%\bea
%i\slashed\partial\,\Phi=2i\left(\begin{matrix} 0 & \partial_+ \\  -\partial_-\,\, & 0 \end{matrix} \right)\Phi=H\,f\,\Phi\equiv  S\Phi
%\label{dirac-lc}
%\ea  

Before considering constant mean curvature surfaces, we give the geometric interpretation of  the corresponding canonical energy momentum tensor of the Dirac system. In the $t, x$ basis:
\bea
&T^{\mu\nu}=\frac{1}{2}
\begin{pmatrix}   
S\, \bar\psi\psi-(Q^{(+)}+Q^{(-)})  &  -(Q^{(+)}-Q^{(-)})  \cr
(Q^{(+)}-Q^{(-)})  &   S\, \bar\psi\psi +(Q^{(+)}+Q^{(-)}) 
 \end{pmatrix} 
% =\frac{i}{2}(\bar\psi\psi)^2\Omega_I^{-1}\,\Omega_{II}& 
\label{emom-1}
\ea
%\bea
%&T^{\mu\nu}=\frac{1}{2}
%\begin{pmatrix}   
%S\, \bar\psi\psi+h_1-h_2  &  h_1+h_2  \cr
% -(h_1+h_2)  &   S\, \bar\psi\psi-(h_1-h_2) 
% \end{pmatrix} 
%% =\frac{i}{2}(\bar\psi\psi)^2\Omega_I^{-1}\,\Omega_{II}& 
%\label{emom-1}
%\ea
%Again the factor $i$ is just an artifact of the opposite signatures of the surface and Dirac system and has no physical significance. 
In terms of the lightcone coordinates, $x_\pm$, the energy-momentum tensor components are related to the Hopf differentials, $Q^{(+)}$ and $Q^{(-)}$, and the mean curvature $H$: 
\bea 
T^{+-}=Q^{(-)}
\quad, \quad
T^{-+}=Q^{(+)}
\quad, \quad
T^{++}=T^{--}=\frac{H}{2}\left(\bar\psi\psi \right)^2
\label{emom-2}
\ea
The energy momentum flow is  identical with the Codazzi equations:
\bea
\partial_\mu T^{\mu\nu}=(\bar\psi\psi)^2\partial^{\nu}S
\quad \leftrightarrow\quad 
\begin{cases}
Q^{(+)}_{-}=\frac{1}{2} f^2 H_+ \cr
%\quad,\quad  
Q^{(-)}_{+}=\frac{1}{2} f^2 H_-
\end{cases}
\ea
Hence the Minkowski version of the surface/Gross-Neveu dictionary is:
\bea
\text{Dirac equation}&:& \, (i\slashed\partial-S)\psi=0 \nn
\text{mean curvature}&:&\,H=S\,(\bar\psi\psi)^{-1}\nn
\text{induced metric factor}&:&\, f=\det\Phi=\bar\psi\psi \nn
\text{Hopf differentials}&:&\,\begin{cases}
Q^{(+)}=-i(\psi^*_1\psi_{1,+}-\psi^*_{1,+}\psi_1)\cr
Q^{(-)}=i(\psi^*_2\psi_{2,-}-\psi^*_{2,-}\psi_2)
\end{cases}
\label{dic_mink}
\ea
[In \cite{Klotzek:2010gp}, the notation $h_1$ and $h_2$ was used for these Hopf differentials: $h_1=-Q^{(+)}$, and $h_2=Q^{(-)}$.]

Minimal surfaces (with $H=0$) are described by a \textit{free massless} Dirac equation, and the corresponding spinors are right/left movers. Constant mean curvature surfaces (with $H=l$) are described  by a \textit{nonlinear} Dirac equation, since $H=l$ implies $S= l\, \bar\psi \psi$:
\bea
\text{Minkowski constant mean curvature surfaces}\quad\leftrightarrow\quad (i\slashed\partial-l\, \bar\psi\psi)\psi=0
\ea
Furthermore, the Gauss equation (\ref{gauss-m}) becomes the ShG equation
\bea
\text{Gauss equation}\quad \leftrightarrow\quad \text{Sinh-Gordon equation}:\quad
\theta_{+-}-2 \sqrt{Q^{(+)}Q^{(-)}}\, l\,\sinh(\theta)=0
\ea
where the metric factor $f^2$ is written as $f^2=\frac{2\sqrt{Q^{(+)}Q^{(-)}}}{l}\, e^\theta$.  In terms of the function $S(x_+, x_-)$ appearing in the Dirac equation, this ShG equation reads:
\bea
S\, S_{+-}-S_+\, S_{-}-\frac{S^4}{4}=-l^2 Q^{(+)}Q^{(-)}
\label{seq-m}
\ea
Note the different signs relative to the Euclidean version (\ref{seq-e}).

\subsection{Minkowski Signature:  Immersion of 2d Surfaces in $AdS_3$}

Our discussion so far gives a geometrical explanation of why the ``Type I'' solutions to the time-dependent Hartree-Fock problem must satisfy the Sinh-Gordon equation, as was argued in   \cite{Klotzek:2010gp,Fitzner:2010nv}. A further observation of  \cite{Klotzek:2010gp,Fitzner:2010nv}  was that these solutions can be mapped directly to classical string solutions in $AdS_3$.  We now show that this follows from our geometrical argument because
one can equivalently view the constant mean curvature surfaces in $\mathbb R^{1,2}$ as minimal surfaces  ($H=0$) in $AdS_3$. Furthermore, classical string solutions  in $AdS_3$  are,  by definition, minimal surfaces. 
For completeness, we give the details of this construction. We first describe the Gauss-Codazzi equations for a general conformal embedding $\R^{1,1}\mapsto AdS_3$, and then the spinor formalism for \textit{minimal} surfaces that correspond to classical string solutions. 

\subsubsection{Basic Differential Geometry of Minkowski Surface Embedding in $AdS_3$}

$AdS_3$ space can be can be realized as a hypersurface in $\R^{2,2}$:
\bea
\vec X.\vec X\equiv -X^2_{-1}-X^2_0+X^2_1+X^2_2=-R^2
\label{ads_norm}
\ea
where $R$ is the AdS radius.
Therefore, the embedding $\R^{1,1}\mapsto AdS_3$ can be formulated naturally as an embedding $\R^{1,1}\mapsto \R^{2,2}$ with the condition (\ref{ads_norm}). In light-cone coordinates, where the induced metric is $ds^2=\frac{2Q^{(+)}Q^{(-)}R^2}{f^2} dx_+\,dx_-$, the moving frame is constructed from the tangent vectors $\vec X_{\pm}$ and two mutually orthogonal normal vectors, $\vec X$ and $\vec N$. The elements of the first and second fundamental forms, $\Omega_I$ and $\Omega_{II}$, are given by the inner products:
\bea
&\vec X_+.\vec X_-=\,\frac{2Q^{(+)}Q^{(-)}R^2}{f^2}
\quad,\quad
\vec X_+.\vec X_+=\vec X_-.\vec X_-=0 &\nn
&\vec X_{++}.\vec N=-iQ^{(+)}
\quad,\quad
\vec X_{++}.\vec N=-iQ^{(-)}
\quad,\quad
\vec X_{+-}.\vec N=\frac{2Q^{(+)}Q^{(-)}\;R^2\,H}{f^2}&
\label{inner-ads}
\ea 
Notice that we have chosen the metric factor differently than the previous cases. The reason for this choice is to make the mapping from minimal surface Gauss-Codazzi equations to the Lax equation explicit. With the same spirit of the previous sections, from the commutativity of second derivatives of the basis vectors, $\vec e_{(i),+-}=\vec e_{(i),-+}$ we obtain the Gauss-Codazzi equations:
\bea
&f\,f_{+-}-f_+\,f_- -\frac{f^4}{4}-f^2\left(\frac{(Q^{(+)}Q^{(-)})_+}{Q^{(+)}Q^{(-)}}\right)_-=-Q^{(+)}Q^{(-)}R^2\left(\frac{1}{R^2}+H^2\right)& \nn
&Q^{(+)}_-=\frac{2Q^{(+)}Q^{(-)}R^2}{f^2} \,H_+&\nn
&Q^{(-)}_+=\frac{2Q^{(+)}Q^{(-)}R^2}{ f^2}\, H_- &
\label{gc-ads}
\ea
When the classical equations of motions 
\bea
\vec X_{+-}-\frac{\vec X_+.\vec X_-}{R^2}\vec X=0
\label{string-eom}
\ea
of the string are imposed we have $H=0$, since $\vec X_{+-}$ is aligned in the direction of $\vec X$. Therefore, a classical worldsheet  of a string in $AdS_3$ is a a minimal surface in the $AdS_3$ space. Once again since $Q^{(+)}$ and $Q^{(-)}$ are right/left moving, with an appropriate isometry we can set them to constants. With this choice, the first of the Gauss-Codazzi equations gives the ShG equation:
\bea
f\,f_{+-}-f_+\,f_- -\frac{f^4}{4 R^2}=-Q^{(+)}Q^{(-)}=\text{constant}
%&\theta_{+-}-\sqrt{-4h_1h_2l^2}\,\sinh\theta=0&
%&\text{where\,\,}f=(-4h_1 h_2 l^2)^{\frac{1}{4}}e^{\frac{\theta}{2}}&\nonumber
\label{sg-ads}
\ea
which can be written as
\bea
\theta_{+-}-2\frac{\sqrt{Q^{(+)}Q^{(-)}}}{R}\, \,\sinh\theta=0
%&\text{where\,\,}f=(-4h_1 h_2 l^2)^{\frac{1}{4}}e^{\frac{\theta}{2}}&\nonumber
\label{sg-ads2}
\ea
where the metric factor is $f^2=2\sqrt{Q^{(+)}Q^{(-)}}\,R\, e^\theta$.

Comparing with the $\R^{1,1}\mapsto\R^{1,2}$ embeddings with constant mean curvature from the previous section, we observe a remarkable duality between the \textit{minimal} surfaces embedded into $AdS_3$, with Gaussian curvature $K=2f^2\ln(f^2)_{+-}$, and \textit{constant mean curvature} surfaces embedded into $\R^{1,2}$, with Gaussian curvature $K=-2f^{-2}\ln(f^2)_{+-}$. They are both characterized by the solutions of the Sinh-Gordon equation. Geometrically, it is a tradeoff between the constant intrinsic curvature $\frac{1}{R}$ of $AdS_3$, and the constant extrinsic curvature $H$.

\subsubsection{Spinor Representation of Minkowski Surfaces in $AdS_3$}

$AdS_3$ is associated with the group $SO(2,2)$, which can be factorized as $SO(2,2)=PSU(1,1) \times PSU(1,1)$. It means that the embedding can be formulated by two spinors as opposed to one. This is consistent with the generalization of the Weierstrass representation for embeddings of surfaces into higher dimensional spaces, as discussed in \cite{konopelchenko-higher,taimanov-higher}. Following the notation of \cite{Klotzek:2010gp}, let us work with two complex coordinates, $Z_1=X_{-1}+iX_{0}$, and  $Z_2=X_{1}+iX_{2}$, normalized as: $-Z_1^*Z_1+Z_2^*Z_2=-R^2$. We introduce two sets of spinors, $\psi_a$ and $\psi_b$, simultaneously  normalized $\bar\psi_a \psi_a=\bar\psi_b\psi_b\equiv\bar\psi\psi$. The spinor parameterization of the surface is given by:
\bea
Z_1=R\, \frac{\bar\psi_a\psi_b}{\bar\psi\psi}\qquad , \qquad Z_2=-R\,\frac{\bar\psi_a i\gamma^5 \psi_b^*}{\bar\psi\psi} 
\ea
Then the Gauss-Codazzi equations (or, equivalently, the classical string equations of motion plus the Virasoro constraints) translate into two Lax equations:
\bea
&\psi_{a,+}=U_a\,\psi_a\quad,\quad\psi_{a,-}=V_a\,\psi_a&\nn
&\psi_{b,+}=U_b\,\psi_b\quad,\quad\psi_{b,-}=V_b\,\psi_b&
\ea 
They also fix the normalization to be $\,\bar\psi\psi=f$.
\bea
\,\bar\psi_a\psi_a=\,\bar\psi_b\psi_b=f\equiv\frac{S}{l}
\ea
Here we identified the spectral parameter l with the $AdS$ scale; $l=\frac{1}{R}$. The Lax pair matrices with this fixed normalization are 
\bea
U_a=\begin{pmatrix} 
\frac{S_+}{S}\, & \frac{iQ^{(+)}\,l}{S}  \cr  -\frac{iS}{2}\,\, & 0 \end{pmatrix} 
\qquad,\qquad 
V_a=\begin{pmatrix} 0\, & \frac{iS}{2} \cr  -\frac{iQ^{(-)}\, l}{S} \, & \frac{S_-}{S} \end{pmatrix} \nn
U_b=\begin{pmatrix} \frac{S_+}{S}\, & \,\,-\frac{iQ^{(+)}\, l}{S}\,\,  \cr -\frac{iS}{2}\,\, & \,\,0\,\, \end{pmatrix} 
 \qquad,\qquad V_b=\begin{pmatrix} 0\, & \frac{if}{2} \cr \,\,\frac{iQ^{(-)}\, l}{S}\,\, \, & \frac{S_-}{S} \end{pmatrix} 
\ea
Consistency of the linear equations gives the zero-curvature conditions
\bea
U_{a,-}-V_{a,+}+[U_a,V_a]=0\qquad , \qquad 
U_{b,-}-V_{b,+}+[U_b,V_b]=0
\label{zc-2}
\ea
and these separately lead to the sinh-gordon equation (\ref{sg-ads}).
Physically, this Dirac system corresponds to the Gross-Neveu model since the "potential" is identified with the scalar interaction term $S=l\,\bar\psi\psi$. 

\section{Static Reduction and Immersions of Curves}

All the discussion so far has been searching for time-dependent solutions $S(x,t)$ to the nonlinear Dirac equation (\ref{nld}). We have seen that $S(x,t)$ must satisfy the ShG equation in the Gauss equation form [we consider here just the Minkowski case, with embedding into ${\mathbb R}^{1,2}$]:
\bea
S\, S_{+-}-S_+\,S_--\frac{1}{4} S^4=-l^2\, Q^{(+)}Q^{(-)}
\label{shg-S}
\ea
To make connection with the previously found static solutions, $S=S(x)$, to the gap equation, based on the NLSE (\ref{nlse}), we note that in light-cone coordinates, assuming simple time dependence of the form $e^{-i E t}$, we can replace:
\bea
\partial_+\to \frac{1}{2}(\partial_x-i\, E)\qquad , \qquad \partial_-\to \frac{1}{2}(\partial_x+i\, E)
\label{der}
\ea
Thus, the Gauss equation (\ref{gauss-m}) becomes
\bea
S\, S^{\prime\prime}-(S^\prime)^2- S^4=-4\,l^2\, Q^{(+)}Q^{(-)}
\label{shg1}
\ea
Now note that any solution of the NLSE (\ref{nlse}), $S^{\prime\prime}-2S^3+c S=0$, has a first integral $(S^\prime)^2=S^4-cS^2+d$, where $c$ and $d$ are constants. For example, the kink crystal solution (\ref{sn}) satisfies the NLSE: $S^{\prime\prime}-2S^3+m^2(1+k^2)S=0$, and also $(S^\prime)^2=S^4-m^2(1+k^2)S^2+m^4k^2$. Thus, any solution to the NLSE satisfies the ShG equation:
\bea
S\, S^{\prime\prime}-(S^\prime)^2-S^4 =-c
\label{shg2}
\ea
which is precisely the Gauss equation form of the ShG equation (\ref{shg1}).
This explains the connection between the static condensates found previously in \cite{Basar:2008ki} and the construction described here. It also explains the similarity between the generalized Weierstrass representation of the embedding of surfaces into $\R^3$ and the Da Rios representation \cite{darios} of the embedding of curves into $\R^3$, also known as the Hasimoto map \cite{hasimoto} in the theory of the motion of vortex filaments. There is a vast literature on the embedding of curves into 3d spaces, and it can be understood as a dimensional reduction of the surface embedding construction \cite{konopelchenko1}.

An important physical observation is that in studying the thermodynamics of the static condensates, we actually need to know the corresponding fermionic energy spectrum, which translates geometrically into studying spectral deformations of the curves that preserve the form of the curve. It is known that the deformations of the NLSE are integrable and are governed by the mKdV integrable hierarchy, and this explains why the thermodynamical Ginzburg-Landau expansion of the grand potential is an expansion in terms of the mKdV conserved quantities, as argued in \cite{Basar:2008ki,Correa:2009xa}. Indeed, if we re-write the Dirac equation (\ref{dirac-m}) with the replacements (\ref{der}), then we obtain a Dirac equation
\bea
\begin{pmatrix}
i\partial_x & S\cr
S &-i\partial_x
\end{pmatrix}
\begin{pmatrix}
\psi_1\cr
\psi_2
\end{pmatrix}=E 
\begin{pmatrix}
\psi_1\cr
\psi_2
\end{pmatrix}
\ea
which is precisely the Bogoliubov-De Gennes equation for the thermodynamics of the system with Dirac equation (\ref{dirac-m}). Geometrically, $E$ is the deformation parameter generating the flow of isometric curves. In Section \ref{GN3} we apply this observation to the Gross-Neveu model in $2+1$ dimensional spacetime.

\subsection{"Boosted" static potentials}
Given a static solution $S=\mathcal S(x)$, it is always possible to generate a class of time-dependent solutions $S(x, t)$ by a Lorentz boost. 
Thus, in the case of Minkowski embedding, given a solution $\mathcal S(x)$ to the static equation (\ref{shg2}), with $c$ rescaled to 1, we can define a new function 
\bea
S(x_+, x_-)=\mathcal S\left(\alpha \, x_+ +\frac{1}{\alpha}\, x_-\right)
\ea
which satisfies the (rescaled) time-dependent Sinh-Gordon equation 
\bea
S\, S_{+-}-S_+\,S_-- S^4=-1
\label{shg-m1}
\ea
Writing $\alpha=e^{-\eta}$, this is a boosted solution with velocity $v=\tanh \eta$:
\bea
S(x_+, x_-)=\mathcal S\left(\frac{2}{\sqrt{1-v^2}}(x-v\, t)\right)
\ea
The spinor appropriately transforms as $\Psi\rightarrow e^{\sigma_3 \eta/2} \Psi$. 

Similarly, for the Euclidean embedding into $\mathbb R^3$, the resulting one-dimensional equation, after rescaling, is
\bea
\mathcal S \mathcal S^{\prime\prime}-(\mathcal S^\prime)^2+\mathcal S^4=1
\ea
Given a solution to this one-dimensional equation, we obtain a two-dimensional solution
\bea
S(z, \bar z)=\mathcal S\left(a \, z +\frac{1}{a}\, \bar z \right)
\ea
which satisfies the (rescaled) time-dependent Sinh-Gordon equation 
\bea
S\, S_{z \bar z}-S_z\,S_{\bar z} +  S^4=1
\label{shg-m2}
\ea
Here the boost acts as a rotation. 

\subsection{Examples}
At this section we recapitulate some known classical string solutions. It is useful to introduce the global coordinates $(t,\rho,\phi)$:
\bea
Z_1=\cosh\rho\, e^{i\,t}\quad\quad Z_2=\sinh\rho \, e^{i\,\phi}
\ea
We also label the worldsheet coordinates as $(\tau,\sigma)$ where the worldsheet metric is conformally $-d\tau^2+d\sigma^2$.

\subsubsection{Spinning string}
The classical solution of the Gubser-Klebanov-Polyakov string is given by
\bea
\rho=\rho(\sigma)\quad\quad t=\kappa\tau\quad\quad \phi=\omega\tau
\ea
with the metric factor
\bea
\rho^{\prime\,2}=\frac{4Q^{(+)}Q^{(-)}}{S^2}=\kappa^2\,\text{sn}^2(\omega\sigma;k^2)\equiv\kappa\omega\,e^{-\theta}
\ea
where $^\prime$ means $\partial_\sigma$. Here the string is parameterized by the elliptic parameter $k^2$ that controls the length (hence the energy and angular momentum) of the string $\omega=\frac{2 \mathbb K(k^2)}{\pi}$, and $\kappa=k\omega$.  The solution satisfies the one dimensional, static sinh-gordon equation:
\bea
&S S^{\prime\prime}-S^{\prime\,2}-S^4=-\frac{\kappa^2\omega^2}{4}&\nn
&\frac{1}{4}\theta^{\prime\prime}=\theta_{+-}=\omega\kappa\sinh\theta&
\ea
where $4 Q^{(+)}Q^{(-)}=\omega^2\kappa^2$.
On the spinor side, it is identified with a fermion system subject to a periodic potential $S=\omega\,\text{ns}(\omega \sigma;k^2)$.

\subsubsection{Pulsating string}
The pulsating string 
\bea
\rho=\rho(\tau)\quad\quad t=\kappa\tau\quad\quad \phi=m\sigma
\ea
 is described by the worldsheet with the metric factor $S^2=-(1-k^2) \omega ^2 \text{nc}^2(\omega \tau;k^2)$
and is associated with the one dimensional, time dependent sinh-gordon equation:
\bea
-\ddot S+\dot S^2-S^4= k^2 (1- k^2) \omega^4
\ea

\section{Implications for Higher-Dimensional Gross-Neveu Models}
\label{GN3}

In $2+1$ dimensional spacetime the Gross-Neveu model is still renormalizable in the large $N_f$ limit \cite{Rosenstein:1990nm}. The phase diagram has been studied via a  gap equation and via Hartree-Fock assuming a uniform condensate \cite{Klimenko:1991he,urlichs}. It has also been studied on the lattice assuming a uniform condensate \cite{Hands:1995jq}. It is still unclear whether there could be an inhomogeneous (static) crystalline condensate that has lower grand potential in some region of the temperature and chemical potential plane. This question was studied by Urlichs \cite{urlichs}, who argued that a stripe condensate of the form (\ref{sn}) satisfies the gap equation but is not thermodynamically preferred. So far, no truly 2d inhomogeneous condensate has been found that satisfies the gap equation. We argue here that if such a condensate is of ``Type I'', then it should satisfy a ShG-type equation.

The Lagrangian of the $2+1$ dimensional Gross-Neveu (${\rm GN}_3$) can be written
\begin{eqnarray}
{\mathcal L}_{\rm GN3}&=&i\,\bar{\Psi}\,\partial\hskip -6pt /\,\Psi +\frac{g^2}{2}\, (\bar \Psi\,\tau_3\,\Psi)^2 
%{\mathcal L}_{\rm NJL}&=&i\,\bar{\psi}\,\partial\hskip -6pt /\,\psi +g\left[(\bar \psi\,\psi)^2+(\bar \psi\,i\, \gamma^5\psi)^2\right]
\label{lag3}
\end{eqnarray}
where we choose the parity-preserving form of the interaction. Here, $\tau_3=\begin{pmatrix} 1 &0\cr 0&-1\end{pmatrix}$, is a color matrix, so that $\Psi$ is in fact a 4-component spinor. The associated Dirac equation is
\begin{eqnarray}
\left(i\partial \hskip -6pt /\hskip 3pt -\Delta\right)\Psi=0 
\label{dirac-3}
\end{eqnarray}
where the parity-even condensate is $\Delta=\bar\Psi \tau_3\Psi$. Since $\tau_3$ is diagonal, we can choose to work with 2-component spinors, $\psi$, with both signs of a condensate $S=\bar\psi\psi$. A suitable choice of Dirac matrices in $2+1$ dimensions is: $\gamma^0=\sigma_3$, $\gamma^1=i\sigma_1$, and $\gamma^2=i\sigma_2$. To seek static solutions, we use the complex coordinates $z=x+iy$ and $\bar z=x-i y$ for the spatial directions, so that the 2-component Dirac equation becomes
\bea
\begin{pmatrix}
S & 2\partial_z\cr
-2\partial_{\bar z} & -S 
\end{pmatrix}
\begin{pmatrix}
\psi_1\cr
\psi_2
\end{pmatrix}=E 
\begin{pmatrix}
\psi_1\cr
\psi_2
\end{pmatrix}
\label{3-deform}
\ea
Again, the eigenvalue $E$ will describe deformations, so to find the basic solution as an embedding of a surface into a 3d space, we first consider the $E=0$ equation. Now, observe that  if $\psi=(\psi_1, \psi_2)^T$ is a solution, then so is $\psi=(\psi_2^*, \psi_1^*)^T$, for $S$ real. Thus, we form the $2\times 2$ matrix $\Phi$ as before:
\bea
\Phi=\begin{pmatrix}
\psi_1 & \psi_2^*\cr
\psi_2 &\psi_1^*
\end{pmatrix}
\ea
Note that 
\bea
\det\Phi=(|\psi_1|^2-|\psi_2|^2) = \bar\psi\psi
\label{3}
\ea
recalling that $\gamma^0=\sigma_3$ in this representation. Thus, if we define the two dimensional static condensate $S(z, \bar z)= \bar\psi\psi$, then we find that this condensate must satisfy the Sinh-Gordon equation:
\bea
S\, S_{z\bar z}-S_z\, S_{\bar z} -\frac{1}{4} S^4 =-|Q|^2
\label{gn3-shg}
\ea
where $Q=(\psi_1\,\psi^*_{2, z}-\psi_2^*\, \psi_{1, z})$. Note that the signs in (\ref{gn3-shg}) are different from those appearing in the Euclidean embedding case (\ref{seq-e}), but the same as those appearing in the Minkowski embedding case (\ref{seq-m}).

This construction is the same form as for the embedding of a 2d surface into the 4d space $\mathbb R^{2,2}$ considered by Konopelchenko and Taimanov \cite{konopelchenko-higher,taimanov-higher}, where  two copies of a $2\times 2$ Dirac equation are needed. Physically, in our case of the ${\rm GN}_3$ model, these two copies
correspond to the two copies making up the full 4-component spinor $\Psi$. So, for a constant mean curvature embedding of ${\mathbb R}^2$ into ${\mathbb R}^{2,1}$ we have a nonlinear Dirac equation of precisely the correct form for a ``Type I'' solution to the ${\rm GN}_3$ gap equation. This static condensate $S$ must satisfy the Gauss equation in its ShG form (\ref{gn3-shg}). Furthermore, to study the thermodynamics of such a  solution we need spectral deformations of this equation, which are governed by the modified Novikov-Veselov (mNV) system \cite{novikov,bogdanov}, which generalizes the mKdV system that governed the thermodynamics of the static condensates in the $1+1$ dimensional Gross-Neveu model  \cite{Basar:2008ki,Correa:2009xa}. 
The thermodynamic implications of this new geometrical perspective on the ${\rm GN}_3$ gap equation are currently being investigated. The step from 1d to 2d is a large one, as is familiar from the intricacies of the inverse scattering method, but our result suggests that if progress can be made in the $2+1$ dimensional system, then the modified Novikov-Veselov system, which is a special case of the Davey-Stewartson system, should play a significant role \cite{novikov,bogdanov,taimanov-mvn}.

\section{Conclusions}

To conclude, we have shown that there is a natural geometric interpretation of the recent results of Thies et al \cite{Klotzek:2010gp,Fitzner:2010nv} concerning time-dependent condensate solutions of the Hartree-Fock problem for the Gross-Neveu model ${\rm GN}_2$. The geometric interpretation is based on the fact that the problem of embedding a 2d surface of constant mean curvature into a 3d space can be expressed as a nonlinear Dirac equation \cite{konopelchenko1,bobenko-integrable,fokas,taimanov-dirac,Novikov:2006zz}, and with the appropriate choice of signature of the surface and the embedding space, this nonlinear Dirac equation is precisely the nonlinear Dirac equation for mode-by-mode solutions of the ${\rm GN}_2$ Hartree-Fock equations. 
This perspective is strongly motivated by earlier work of Regge and Lund \cite{Lund:1976ze}, Pohlmeyer \cite{Pohlmeyer:1975nb}, and Neveu and Papanicolaou \cite{Neveu:1977cr}. 
For these Gross-Neveu solutions, the logarithm of the bilinear condensate automatically satisfies the Sinh-Gordon equation, which geometrically is the Gauss equation for the induced metric factor.  The connection with classical string solutions observed in \cite{Klotzek:2010gp,Fitzner:2010nv} follows immediately once we identify a constant mean curvature solution in $\mathbb R^{1,2}$ with a  zero mean curvature surface (i.e., a classical string solution) embedded in a 3d space of constant negative curvature, ${\rm AdS}_3$. The relation to the explicit string solutions in  \cite{Jevicki:2007aa} follows from the fundamental role of the Sinh-Gordon equation in the construction of string solutions \cite{Pohlmeyer:1975nb,Lund:1976ze,Neveu:1977cr,Jevicki:2007aa,DeVega:1992xc,Grigoriev:2007bu}. This geometrical perspective on the Gross-Neveu gap equation and Hartree-Fock equations permits us to address the currently unsolved question of the possibility of inhomogeneous condensates in the $2+1$ dimensional Gross-Neveu model. We find that a static condensate that satisfies the Hartree-Fock equations in the same mode-by-mode manner, must satisfy a Sinh-Gordon equation. The thermodynamical implications of this result are currently being investigated.

We have not yet addressed the so-called ``Type II'' solutions of the {\it massive} ${\rm GN}_2$ system or of the massless ${\rm NJL}_2$ system, each of which satisfies a generalized nonlinear Dirac equation. Assuming a nonlinear Dirac equation, with a mode-by-mode solution (i.e. a ``Type I'' solution), the ${\rm NJL}_2$ system leads to a clear geometrical interpretation \cite{Pohlmeyer:1975nb,Lund:1976ze,Neveu:1977cr}. But it is known that the static condensates for this system are not necessarily mode-by-mode solutions of the nonlinear Dirac equation, except for the simple chiral spiral solution $\Delta(x)=A e^{i\mu x}$ (which, incidentally,  is the thermodynamically preferred solution of the massless ${\rm NJL}_2$ system). These more general condensates satisfy more complicated nonlinear equations, and it would be interesting to investigate their possible geometric form, along with the connection to brane-world scenarios of Gross-Neveu models \cite{Bietenholz:2003wa}.
We also note that nonlinear Dirac equations have been used recently in the analysis of Bose-Einstein condensates in a two-dimensional  honeycomb lattice \cite{carr}. It would be interesting to see if this geometrical approach could be useful in such systems. Another curious relation worth mentioning is the fact that the Gauss-Codazzi equations are also naturally related to the Hitchin equations \cite{Ward:1985gz,Hitchin:1986vp}, which in turn arise in the relation between self-dual Chern-Simons solitons  and integrable models  \cite{Grossman:1990it,Dunne:1990qe,Dunne:1995ai,soliani}. This is a reflection of the sigma model form of the Pohlmeyer reduction problem \cite{Grigoriev:2007bu}.

\section{Acknowledgments}
We thank the DOE for support under grant DE-FG02-92ER40716, and M. Thies for helpful correspondence.


\begin{thebibliography}{99}

%\cite{Gross:1974jv}
\bibitem{Gross:1974jv}
  D.~J.~Gross and A.~Neveu,
  ``Dynamical Symmetry Breaking In Asymptotically Free Field Theories,''
  Phys.\ Rev.\  D {\bf 10}, 3235 (1974).
  %%CITATION = PHRVA,D10,3235;%%

%\cite{Dashen:1975xh}
\bibitem{Dashen:1975xh}
  R.~F.~Dashen, B.~Hasslacher and A.~Neveu,
  ``Semiclassical Bound States In An Asymptotically Free Theory,''
  Phys.\ Rev.\  D {\bf 12}, 2443 (1975).
  %%CITATION = PHRVA,D12,2443;%%

%\cite{Feinberg:2003qz}
\bibitem{Feinberg:2003qz}
  J.~Feinberg,
  ``All about the static fermion bags in the Gross-Neveu model,''
  Annals Phys.\  {\bf 309}, 166 (2004)
  [arXiv:hep-th/0305240].
  %%CITATION = APNYA,309,166;%%
  
   
    %\cite{Thies:2003kk}
\bibitem{Thies:2003kk}
  M.~Thies and K.~Urlichs,
  ``Revised phase diagram of the Gross-Neveu model,''
  Phys.\ Rev.\  D {\bf 67}, 125015 (2003)
  [arXiv:hep-th/0302092];
  %%CITATION = PHRVA,D67,125015;%%
M.~Thies,
  ``Analytical solution of the Gross-Neveu model at finite density,''
  Phys.\ Rev.\  D {\bf 69}, 067703 (2004)
  [arXiv:hep-th/0308164].
  %%CITATION = PHRVA,D69,067703;%%
     
  %\cite{Thies:2006ti}
\bibitem{Thies:2006ti}
  M.~Thies,
  ``From relativistic quantum fields to condensed matter and back again:
  Updating the Gross-Neveu phase diagram,''
  J.\ Phys.\ A  {\bf 39}, 12707 (2006)
  [arXiv:hep-th/0601049].
  %%CITATION = JPAGB,A39,12707;%%
   
  %\cite{Schon:2000he}
\bibitem{Schon:2000he}
  V.~Schon and M.~Thies,
  ``Emergence of Skyrme crystal in Gross-Neveu and 't Hooft models at  finite
  density,''
  Phys.\ Rev.\  D {\bf 62}, 096002 (2000)
  [arXiv:hep-th/0003195];
  %%CITATION = PHRVA,D62,096002;%%
  V.~Schon and M.~Thies,
  ``2D model field theories at finite temperature and density,''
  arXiv:hep-th/0008175,
  in {\it At the frontier of particle physics, vol. 3}, page 1945, 
Festschrift in honor of Boris Ioffe, edited by M. Shifman (World Scientific, 2000).
  %%CITATION = HEP-TH/0008175;%%
  
   %\cite{Basar:2009fg}
\bibitem{Basar:2009fg}
  G.~Basar, G.~V.~Dunne and M.~Thies,
  ``Inhomogeneous Condensates in the Thermodynamics of the Chiral NJL2
  model,''
  Phys.\ Rev.\  D {\bf 79}, 105012 (2009)
  [arXiv:0903.1868 [hep-th]].
  %%CITATION = PHRVA,D79,105012;%%
  
    %\cite{deForcrand:2006zz}
\bibitem{deForcrand:2006zz}
  P.~de Forcrand and U.~Wenger,
  ``New baryon matter in the lattice Gross-Neveu model,''
  PoS {\bf LAT2006}, 152 (2006)
  [arXiv:hep-lat/0610117].
  %%CITATION = POSCI,LAT2006,152;%%
  
  %\cite{Nickel:2008ng}
\bibitem{Nickel:2008ng}
  D.~Nickel and M.~Buballa,
  ``Solitonic ground states in (color-) superconductivity,''
  Phys.\ Rev.\  D {\bf 79}, 054009 (2009)
  [arXiv:0811.2400 [hep-ph]];
  %%CITATION = PHRVA,D79,054009;%%
D.~Nickel,
  ``How many phases meet at the chiral critical point?,''
  Phys.\ Rev.\ Lett.\  {\bf 103}, 072301 (2009)
  [arXiv:0902.1778 [hep-ph]];
  %%CITATION = PRLTA,103,072301;%%
%\cite{Nickel:2009wj}
  ``Inhomogeneous phases in the Nambu-Jona-Lasino and quark-meson model,''
  Phys.\ Rev.\  D {\bf 80}, 074025 (2009)
  [arXiv:0906.5295 [hep-ph]].
  %%CITATION = PHRVA,D80,074025;%%
  
   \bibitem{glozman}
  L.~Y.~Glozman and R.~F.~Wagenbrunn,
  ``Chirally symmetric but confining dense and cold matter,''
  Phys.\ Rev.\  D {\bf 77}, 054027 (2008)
  [arXiv:0709.3080 [hep-ph]];
  %%CITATION = PHRVA,D77,054027;%%
  ``Second order chiral restoration phase transition at low temperatures in quarkyonic matter,''
  arXiv:0805.4799 [hep-ph].
  %%CITATION = ARXIV:0805.4799;%%
  
  %\cite{Kojo:2010zz}
\bibitem{Kojo:2010zz}
  T.~Kojo, Y.~Hidaka, L.~McLerran and R.~D.~Pisarski,
  ``Quarkyonic chiral spirals,''
  AIP Conf.\ Proc.\  {\bf 1257}, 732 (2010);
  %%CITATION = APCPC,1257,732;%%
   T.~Kojo, R.~D.~Pisarski and A.~M.~Tsvelik,
  ``Covering the Fermi Surface with Patches of Quarkyonic Chiral Spirals,''
  Phys.\ Rev.\  D {\bf 82}, 074015 (2010)
  [arXiv:1007.0248 [hep-ph]].
  %%CITATION = PHRVA,D82,074015;%%

%\cite{Frolov:2010wn}
\bibitem{Frolov:2010wn}
  I.~E.~Frolov, V.~C.~Zhukovsky and K.~G.~Klimenko,
  ``Chiral density waves in quark matter within the Nambu--Jona-Lasinio model
  in an external magnetic field,''
  Phys.\ Rev.\  D {\bf 82}, 076002 (2010)
  [arXiv:1007.2984 [hep-ph]].
  %%CITATION = PHRVA,D82,076002;%%



 
  %\cite{Pohlmeyer:1975nb}
\bibitem{Pohlmeyer:1975nb}
  K.~Pohlmeyer,
  ``Integrable Hamiltonian Systems And Interactions Through Quadratic
  Constraints,''
  Commun.\ Math.\ Phys.\  {\bf 46}, 207 (1976).
  %%CITATION = CMPHA,46,207;%%
  
  %\cite{Lund:1976ze}
\bibitem{Lund:1976ze}
  F.~Lund and T.~Regge,
  ``Unified Approach To Strings And Vortices With Soliton Solutions,''
  Phys.\ Rev.\  D {\bf 14}, 1524 (1976);
  %%CITATION = PHRVA,D14,1524;%%
   F.~Lund,
  ``Note On The Geometry Of The Nonlinear Sigma Model In Two-Dimensions,''
  Phys.\ Rev.\  D {\bf 15}, 1540 (1977);
  %%CITATION = PHRVA,D15,1540;%%
  F. Lund,
  ``Solitons and Geometry'',
  In {\it Nonlinear Equations In Physics and Mathematics}, (Proceedings, Istanbul 1977), A. O. Barut (Ed), (D. Reidel, Boston, 1978).  
 
  %\cite{Neveu:1977cr}
\bibitem{Neveu:1977cr}
  A.~Neveu and N.~Papanicolaou,
  ``Integrability Of The Classical Scalar And Symmetric Scalar-Pseudoscalar
  Contact Fermi Interactions In Two-Dimensions,''
  Commun.\ Math.\ Phys.\  {\bf 58}, 31 (1978).
  %%CITATION = CMPHA,58,31;%%
  
  
  

  
  %\cite{Basar:2008ki}
\bibitem{Basar:2008ki}
G.~Basar and G.~V.~Dunne,
  ``Self-consistent crystalline condensate in chiral Gross-Neveu and
  Bogoliubov-de Gennes systems,''
  Phys.\ Rev.\ Lett.\  {\bf 100}, 200404 (2008)
  [arXiv:0803.1501 [hep-th]];
  %%CITATION = PRLTA,100,200404;%%
  G.~Basar and G.~V.~Dunne,
  ``A Twisted Kink Crystal in the Chiral Gross-Neveu model,''
  Phys.\ Rev.\  D {\bf 78}, 065022 (2008)
  [arXiv:0806.2659 [hep-th]].
  %%CITATION = PHRVA,D78,065022;%%
  
     %\cite{Correa:2009xa}
\bibitem{Correa:2009xa}
F.~Correa, G.~V.~Dunne and M.~S.~Plyushchay,
        ``The Bogoliubov/de Gennes system, the AKNS hierarchy, and nonlinear quantum
        mechanical supersymmetry,''
        Annals Phys.\  {\bf 324}, 2522 (2009)
        [arXiv:0904.2768 [hep-th]].
        %%CITATION = APNYA,324,2522;%%
        
\bibitem{gesztesy} 
F.~Gesztesy and H.~Holden, 
  {\it Soliton Equations and their Algebro-Geometric Solutions}, 
  (Cambridge University Press, 2003)

        
         %\cite{Ablowitz:1974ry}
\bibitem{Ablowitz:1974ry}
  M.~J.~Ablowitz, D.~J.~Kaup, A.~C.~Newell and H.~Segur,
  ``The Inverse scattering transform fourier analysis for nonlinear problems,''
  Stud.\ Appl.\ Math.\  {\bf 53}, 249 (1974).
  %%CITATION = SAPMB,53,249;%%


  %\cite{Klotzek:2010gp}
\bibitem{Klotzek:2010gp}
  A.~Klotzek and M.~Thies,
  ``Kink dynamics, sinh-Gordon solitons and strings in AdS(3) from the
  Gross-Neveu model,''
  J.\ Phys.\ A  {\bf 43}, 375401 (2010)
  [arXiv:1006.0324 [hep-th]].
  %%CITATION = JPAGB,A43,375401;%%
  
  %\cite{Fitzner:2010nv}
\bibitem{Fitzner:2010nv}
  C.~Fitzner and M.~Thies,
  ``Exact solution of an N baryon problem in the Gross-Neveu model,''
  arXiv:1010.5322 [hep-th].
  %%CITATION = ARXIV:1010.5322;%%
  
  %\cite{Jevicki:2007aa}
\bibitem{Jevicki:2007aa}
  A.~Jevicki, K.~Jin, C.~Kalousios and A.~Volovich,
  ``Generating AdS String Solutions,''
  JHEP {\bf 0803}, 032 (2008)
  [arXiv:0712.1193 [hep-th]];
  %%CITATION = JHEPA,0803,032;%%
  A.~Jevicki and K.~Jin,
  ``Solitons and AdS String Solutions,''
  Int.\ J.\ Mod.\ Phys.\  A {\bf 23}, 2289 (2008)
  [arXiv:0804.0412 [hep-th]].
  %%CITATION = IMPAE,A23,2289;%%
  
  %\cite{Antonyan:2006qy}
\bibitem{Antonyan:2006qy}
  E.~Antonyan, J.~A.~Harvey and D.~Kutasov,
  ``The Gross-Neveu model from string theory,''
  Nucl.\ Phys.\  B {\bf 776}, 93 (2007)
  [arXiv:hep-th/0608149].
  %%CITATION = NUPHA,B776,93;%%
  
  %\cite{Basu:2006eb}
\bibitem{Basu:2006eb}
  A.~Basu and A.~Maharana,
  ``Generalized Gross-Neveu models and chiral symmetry breaking from string
  theory,''
  Phys.\ Rev.\  D {\bf 75}, 065005 (2007)
  [arXiv:hep-th/0610087].
  %%CITATION = PHRVA,D75,065005;%%

  
  %\cite{Davis:2007ka}
\bibitem{Davis:2007ka}
  J.~L.~Davis, M.~Gutperle, P.~Kraus and I.~Sachs,
  ``Stringy NJL and Gross-Neveu models at finite density and temperature,''
  JHEP {\bf 0710}, 049 (2007)
  [arXiv:0708.0589 [hep-th]].
  %%CITATION = JHEPA,0710,049;%%
  
%  \bibitem{bobenko}
%A. I. Bobenko, 
%``Integrable surfaces'',
%Func. Anal. Appl. {\bf 24}, 227 (1990).

\bibitem{eisenhart-book}
L. P. Eisenhart, 
{\it A Treatise on the Differential Geometry of Curves and Surfaces},
(Ginn Co., Boston, 1909).

\bibitem{hopf-book}
H. Hopf,
{\it Differential Geometry in the Large}, Lect. Notes Math. {\bf 1000}, (Springer, Berlin, 1983).

  \bibitem{kenmotsu}
K. Kenmotsu, ``Weierstrass formula for surfaces of prescribed mean curvature '',  
Math. Ann. {\bf 245}, 89 (1979).

\bibitem{bobenko-integrable}
A. I. Bobenko, 
``Integrable surfaces'',
Func. Anal. Appl. {\bf 24}, 227 (1990);
 ``Constant mean curvature surfaces and integrable equations'', 
Russ. Math. Surv. {\bf 46}, 1 (1991);
``All constant mean curvature tori in R3, S3 and H3 in terms of theta-functions'',
  Math. Ann. {\bf 290}, 209 (1991);
 ``Exploring Surfaces through Methods from the Theory of Integrable Systems:
  Lectures on the Bonnet problem,''
  [arXiv:math/9909003],
 in {\it Surveys on Geometry and Integrable Systems}, 
 Advanced Studies in Pure Mathematics {\bf 51}, 1 (2008).
  %%CITATION = MATH/9909003;%%

  \bibitem{konopelchenko1}
B. G. Konopelchenko,  
``Induced surfaces and their integrable dynamics'',  
Stud. Appl. Math. {\bf 96}, 9 (1996);
B. G. Konopelchenko and I. A. Taimanov, 
``Constant mean curvature surfaces via integrable dynamical system'',
J. Phys. A {\bf 29}, 1261 (1996) [arXiv:dg-ga/9505006]. 
  
  \bibitem{fokas}
A.S. Fokas and I.M. Gel'fand,
``Surfaces on Lie Groups, on Lie Algebras, and Their Integrability'',
Commun. Math. Phys. {\bf 177}, 203  (1996);
A. S. Fokas, I. M. GelÕfand, F. Finkel and Q. M. Liu,
``A formula for constructing inÞnitely many surfaces on Lie algebras and integrable equations'',
Sel. Math., New ser. {\bf 6}, 347 (2000).

\bibitem{taimanov-dirac}
I.~A.~Taimanov,
``Two-dimensional Dirac operator and the theory of surfaces'',
Russ. Math. Surv. {\bf 61}, 79 (2006).

%\cite{Novikov:2006zz}
\bibitem{Novikov:2006zz}
  S.~P.~Novikov and I.~A.~Taimanov,
  {\it Modern geometric structures and fields}, 
%\href{http://www.slac.stanford.edu/spires/find/hep/www?irn=7623089}{SPIRES entry}
(AMS, Providence,  2006).  

\bibitem{darios}
L. S. Da Rios, ``Moto dÕun liquido indefinito con un filetto vorticoso di forma qualunque (On the motion of an unbounded liquid with a vortex filament of any shape)'',  
Rend. Circ. Mat. Palermo {\bf 22}, 117 (1906);
See R. L. Ricca, ``Rediscovery of the Da Rios Equations'', Nature {\bf 352}, 561 (1991).

\bibitem{lamb}
  G.~L.~Lamb,
  ``Analytical Descriptions of Ultrashort Optical Pulse Propagation in a Resonant Medium,''
  Rev.\ Mod.\ Phys.\  {\bf 43}, 99 (1971);
  %%CITATION = RMPHA,43,99;%%
G.~L.~Lamb,
  ``Solitons On Moving Space Curves,''
  J.\ Math.\ Phys.\  {\bf 18}, 1654 (1977);
  %%CITATION = JMAPA,18,1654;%%
  G.~L.~Lamb,
  ``Solitons And The Motion Of Helical Curves,''
  Phys.\ Rev.\ Lett.\  {\bf 37}, 235 (1976).
  %%CITATION = PRLTA,37,235;%%
  
  \bibitem{hasimoto}
R. Hasimoto, 
``A soliton on a vortex filament'', 
J. Fluid Mech. {\bf 51}, 477 (1972).
  
  %\cite{Sym:1979er}
\bibitem{Sym:1979er}
  A.~Sym and J.~Corones,
  ``Lie Group Explanation Of Geometric Interpretations Of Solitons,''
  Phys.\ Rev.\ Lett.\  {\bf 42}, 1099 (1979);
  %%CITATION = PRLTA,42,1099;%%
  A. Sym,
``Soliton Surfaces'',
Lett. Nuovo Cim. {\bf 332}, 394 (1982).

  \bibitem{dodd}
  R. K. Dodd,
  ``Soliton Immersions'',
Commun. Math. Phys. {\bf 197}, 641 (1998).
  
  \bibitem{calini}
  A. Calini and T. Ivey, 
  ``Connecting geometry, topology and spectra for finite-gap NLS potentials'', 
  Physica D {\bf 152}, 9 (2001).
  
  \bibitem{schmidt}
  P. G. Grinevich and M. U. Schmidt, 
  ``Closed curves in R3 : a characterization in terms of curvature and torsion, the Hasimoto map and periodic 
solutions of the filament equation'',  
Preprint: dg-ga/9703020.

 \bibitem{wolff}
U.~Wolff,
``The phase diagram of the infinite-N Gross-Neveu Model at finite temperature and chemical potential'', Phys.\ Lett.\ {\bf 157B}, 303 (1985).

\bibitem{treml}
  T.~F.~Treml,
  ``Dynamical Mass Generation In The Gross-Neveu Model At Finite Temperature And Density,''
  Phys.\ Rev.\  D {\bf 39}, 679 (1989).
  %%CITATION = PHRVA,D39,679;%%
  
  \bibitem{barducci}
  A.~Barducci, R.~Casalbuoni, M.~Modugno, G.~Pettini and R.~Gatto,
  ``Thermodynamics Of The Massive Gross-Neveu Model,''
  Phys.\ Rev.\  D {\bf 51}, 3042 (1995)
  [arXiv:hep-th/9406117].
  %%CITATION = PHRVA,D51,3042;%%

\bibitem{konopelchenko-higher}
B.G. Konopelchenko and G. Landolfi, 
``Generalized Weierstrass representation for surfaces in multidimensional Riemann spaces'',
Journ. Geom. Phys. {\bf 29}, 319 (1999) [arXiv:math/9804144];
B.G. Konopelchenko,
``Weierstrass representations for surfaces in 4D spaces and their integrable deformations via DS hierarchy'', Ann. Global Anal. Geom. {\bf 16}, 61 (2000) [arXiv:math/9807129].

\bibitem{taimanov-higher} 
I. A. Taimanov,
``Surfaces in the four-space and the Davey--Stewartson equations'',
J.  Geometry and Physics {\bf 56}, 1235 (2006), [arXiv:math/0401412]. 

  %\cite{Rosenstein:1990nm}
\bibitem{Rosenstein:1990nm}
  B.~Rosenstein, B.~Warr and S.~H.~Park,
  ``Dynamical symmetry breaking in four Fermi interaction models,''
  Phys.\ Rept.\  {\bf 205}, 59 (1991);
  %%CITATION = PRPLC,205,59;%%
  G.~Gat, A.~Kovner and B.~Rosenstein,
  ``Chiral phase transitions in d = 3 and renormalizability of four Fermi
  interactions,''
  Nucl.\ Phys.\  B {\bf 385}, 76 (1992).
  %%CITATION = NUPHA,B385,76;%%
  
  %\cite{Klimenko:1991he}
\bibitem{Klimenko:1991he}
  K.~G.~Klimenko,
  ``Three-dimensional Gross-Neveu model at nonzero temperature and in an
  external magnetic field,''
  Z.\ Phys.\  C {\bf 54}, 323 (1992).
  %%CITATION = ZEPYA,C54,323;%%


  
  \bibitem{urlichs}
  K. Urlichs, `Baryons and baryonic matter in four-fermon interaction models'', 
  PhD Thesis, Univ. Erlangen, 2007, and unpublished notes (2007).
  
  %\cite{Hands:1995jq}
\bibitem{Hands:1995jq}
  S.~Hands, S.~Kim and J.~B.~Kogut,
  ``The U(1) Gross-Neveu model at nonzero chemical potential,''
  Nucl.\ Phys.\  B {\bf 442}, 364 (1995)
  [arXiv:hep-lat/9501037];
  %%CITATION = NUPHA,B442,364;%%
  S.~Hands,
  ``Four fermion models at non-zero density,''
  Nucl.\ Phys.\  A {\bf 642}, 228 (1998)
  [arXiv:hep-lat/9806022].
  %%CITATION = NUPHA,A642,228;%%
  
    \bibitem{novikov}
S. P. Novikov and A. P. Veselov, 
``Finite-zone two-dimensional periodic Shr¬odinger operators: potential operators'', 
Dokl. Akad. Nauk SSSR {\bf 279}, 784 (1984);
``Two-dimensional Schr¬odinger operator: inverse scattering and evolutional equations'', 
Physica D {\bf 18}, 267 (1986);
``Exactly solvable two-dimensional Schrodinger operators and Laplace transformations'',
{\it Solitons, geometry, and topology: on the crossroad}, 
Amer. Math. Soc. Transl. Ser. 2, {\bf 179}, 109 (1997) [arXiv:math-ph/0003008v1].

\bibitem{bogdanov}
L. V. Bogdanov, 
``Veselov-Novikov equation as a natural two-dimensional generalization of the Korteweg-De Vries equation'',
Theor. Math. Phys. {\bf 70}, 219 (1987);
L. V. Bogdanov, 
``On the two-dimensional Zakharov-Shabat Problem'',
Theor. Math. Phys. {\bf 72}, 790 (1988).

\bibitem{taimanov-mvn}
 I. A. Taimanov,  ``Modified Novikov--Veselov equation and differential geometry of surfaces'', 
Amer. Math. Soc. Transl., Ser. 2, {\bf 179}, 133 (1997) [arXiv:dg-ga/9511005]. 

 \bibitem{DeVega:1992xc}
  H.~J.~De Vega and N.~G.~Sanchez,
  ``Exact Integrability Of Strings In D-Dimensional De Sitter Space-Time,''
  Phys.\ Rev.\  D {\bf 47}, 3394 (1993).
  %%CITATION = PHRVA,D47,3394;%%

  %\cite{Grigoriev:2007bu}
\bibitem{Grigoriev:2007bu}
  M.~Grigoriev and A.~A.~Tseytlin,
  ``Pohlmeyer reduction of $AdS_5 \times S^5$ superstring sigma model,''
  Nucl.\ Phys.\  B {\bf 800}, 450 (2008)
  [arXiv:0711.0155 [hep-th]].
  %%CITATION = NUPHA,B800,450;%%
  
   %\cite{Bietenholz:2003wa}
\bibitem{Bietenholz:2003wa}
  W.~Bietenholz, A.~Gfeller and U.~J.~Wiese,
  ``Dimensional reduction of fermions in brane worlds of the Gross-Neveu
  model,''
  JHEP {\bf 0310}, 018 (2003)
  [arXiv:hep-th/0309162].
  %%CITATION = JHEPA,0310,018;%%


  \bibitem{carr}
L. H. Haddad and L. D. Carr, ``The Nonlinear Dirac Equation in Bose-Einstein Condensates: Relativistic Linear Stability Equations, Nonlinear Localized Modes, and Cherenkov Radiation'', arXiv:1006.3893v2; 
``The Nonlinear Dirac Equation in Bose-Einstein Condensates: Foundation and Symmetries'',
Physica D {\bf 238}, 1413 (2009) [arXiv:0803.3039v1].

  %\cite{Ward:1985gz}
\bibitem{Ward:1985gz}
  R.~S.~Ward,
  ``Integrable and solvable systems, and relations among them,''
  Phil.\ Trans.\ Roy.\ Soc.\ Lond.\  A {\bf 315}, 451 (1985).
  %%CITATION = PTRSA,A315,451;%%
  
  %\cite{Hitchin:1986vp}
\bibitem{Hitchin:1986vp}
  N.~J.~Hitchin,
  ``The Selfduality Equations On A Riemann Surface,''
  Proc.\ Lond.\ Math.\ Soc.\  {\bf 55}, 59 (1987).
  %%CITATION = PLMTA,55,59;%%
  
  %\cite{Grossman:1990it}
\bibitem{Grossman:1990it}
  B.~Grossman,
  ``Hierarchy of soliton solutions to the gauged nonlinear Schrodinger equation
  on the plane,''
  Phys.\ Rev.\ Lett.\  {\bf 65}, 3230 (1990).
  %%CITATION = PRLTA,65,3230;%%

  %\cite{Dunne:1990qe}
\bibitem{Dunne:1990qe}
  G.~V.~Dunne, R.~Jackiw, S.~Y.~Pi and C.~A.~Trugenberger,
  ``Selfdual Chern-Simons solitons and two-dimensional nonlinear equations,''
  Phys.\ Rev.\  D {\bf 43}, 1332 (1991)
  [Erratum-ibid.\  D {\bf 45}, 3012 (1992)]
  [Phys.\ Rev.\  D {\bf 45}, 3012 (1992)].
  %%CITATION = PHRVA,D45,3012;%%
  
  %\cite{Dunne:1995ai}
\bibitem{Dunne:1995ai}
  G.~V.~Dunne,
 {\it Selfdual Chern-Simons Theories}, 
  Lect.\ Notes Phys.\  {\bf M36}, 1 (Springer, Heidelberg, 1995).
  %%CITATION = LNPHA,M36,1;%%
  
  \bibitem{soliani}
L. Martina, Kur. Myrzakul,  R. Myrzakulov and G. Soliani,
``Deformation of surfaces, integrable systems, and Chern-Simons theory'',
Journ. Math. Phys. {\bf 42},  1397 (2001).

%%\cite{Bobenko:1999xb}
%\bibitem{Bobenko:1999xb}
%  A.~I.~Bobenko,
%  ``Exploring Surfaces through Methods from the Theory of Integrable Systems.
%  lectures on the Bonnet problem,''
%  [arXiv:math/9909003],
% in {\it Surveys on Geometry and Integrable Systems}, Advanced Studies in Pure Mathematics {\bf 51}, 1 (2008).
%  %%CITATION = MATH/9909003;%%
  
%  %\cite{Bobenko:1993yp}
%\bibitem{Bobenko:1993yp}
%  A.~I.~Bobenko,
%  ``Surfaces in terms of 2 by 2 matrices. Old and new integrable cases,''
%in {\it Harmonic Maps and Integrable Systems}, ed. by A. Fordy and J. Wood, (Aspects of Mathematics, Vieweg, 1993). 
%  %%CITATION = SFB-288-66;%%

%  \bibitem{taimanov}
%  I. A. Taimanov, 
%   ``The Weierstrass representation of closed surfaces in $R^3$'', 
%Func. Anal. Appl. {\bf 32}, 49 (1998).

%\bibitem{taimanov-mvn}
% I. A. Taimanov,  ``Modified Novikov--Veselov equation and differential geometry of surfaces'', 
%Amer. Math. Soc. Transl., Ser. 2, {\bf 179}, 133 (1997) [arXiv:dg-ga/9511005]. 

%\bibitem{bullough}
% R. K. Dodd and R. K. Bullough,
%``Backlund transformations for the sine-Gordon equations''
%Proc. R. Soc. Lond. A. {\bf 351}, 499 (1976). 
    
%  %\cite{Dunne:1992hq}
%\bibitem{Dunne:1992hq}
%  G.~V.~Dunne,
%  ``Chern-Simons solitons, toda theories and the chiral model,''
%  Commun.\ Math.\ Phys.\  {\bf 150}, 519 (1992)
%  [arXiv:hep-th/9204056].
%  %%CITATION = CMPHA,150,519;%%
  
%  %\cite{Dunne:1998qy}
%\bibitem{Dunne:1998qy}
%  G.~V.~Dunne,
%  ``Aspects of Chern-Simons theory,'' [arXiv:hep-th/9902115],
%  Lectures  at Les Houches Summer School, Session 69: {\it Topological Aspects of Low-dimensional Systems},  July 1998; Proceedings edited by A. Comtet, T. Jolicoeur, S. Ouvry, F. David (Springer- Verlag, Berlin, 2000).
%  %%CITATION = HEP-TH/9902115;%

%\bibitem{bianchi} 
%L. Bianchi, 
%``Ricerche sulle superficie a curvatura constante e sulle elicoidi'', Tesi di Abilitazione, 
%Ann. Scuola Norm. Sup. Pisa {\bf 2}, 285 (1879); 
%``Sopra i sistemi tripli ortogonali di Weingarten'', 
%Ann Matematica {\bf 13}, 177 (1879); 
%``Sopra alcune nuove classi di superficie e di sistemi tripli ortogonali'', 
%Ann. Mat. Pura Appl. {\bf 18}, 301 (1890); 
%{\it Lezioni di Geometria Differenziale}, vol 1 (Pisa: Enrico Spoerri, 1922).

%

%\bibitem{backlund}
%A. V. B\"acklund, 
%``Om ytor med konstant negative kršning'', 
%Lund Universitets Arsskrif {\bf 19} 1, (1883).

%\bibitem{darboux}
%G. Darboux, 
%``Sur les surfaces dont la courbure totale est constante'',  
%Comptes Rendus {\bf 97}, 848, 892, and 946, (1883).

%\bibitem{enneper}
%A. Enneper,  
%``Analytisch-geometrische untersuchungen'',   
%Nachr. konigl. Gesell. Wiss., Georg August. Univ. Gottingen. {\bf 12}, 232 (1868).

%  
%  %\cite{Lund:1977dt}
%\bibitem{Lund:1977dt}
%  F.~Lund,
%  ``Example Of A Relativistic, Completely Integrable, Hamiltonian System,''
%  Phys.\ Rev.\ Lett.\  {\bf 38}, 1175 (1977).
%  %%CITATION = PRLTA,38,1175;%%
%  
%  %\cite{Lund:1977vg}
%\bibitem{Lund:1977vg}
%  F.~Lund,
%  ``Classically Solvable Field Theory Model,''
%  Annals Phys.\  {\bf 115}, 251 (1978).
%  %%CITATION = APNYA,115,251;%%

%  \bibitem{maison}
%J. Math. Phys. 20, 871 (1979); doi:10.1063/1.524134 (7 pages)
%On the complete integrability of the stationary, axially symmetric Einstein equations,
%Dieter Maison 
%


  
    
  
   

  







\end{thebibliography}
\end{document}